\documentclass[hyper]{JHEP3}

\usepackage{epsfig}
\usepackage{latexsym}
\usepackage{amsfonts}
\usepackage{amsmath}
\usepackage{amsthm}
\usepackage{amssymb}
\usepackage{amsbsy}
\usepackage{multirow}

\newcommand{\Dp}[1]{\mathrm{D {#1}}}
\newcommand{\antiDp}[1]{\overline{\Dp{#1}}}
\newcommand{\dihalo}{D6$\antiDp{6}$D0 }

\newcommand{\ova}{ {\overline{a} } }
\newcommand{\GmSix}{\Gamma_6}
\newcommand{\GmSixb}{\Gamma_{\overline{6}}}
\newcommand{\GmA}{\Gamma_a}

\def\calk         {{\cal K}}

\def\caln         {{\cal N}}

\def\calz         {{\cal Z}}
\newsavebox{\uuunit}
\sbox{\uuunit}
    {\setlength{\unitlength}{0.825em}
     \begin{picture}(0.6,0.7)
        \thinlines
        \put(0,0){\line(1,0){0.5}}
        \put(0.15,0){\line(0,1){0.7}}
        \put(0.35,0){\line(0,1){0.8}}
       \multiput(0.3,0.8)(-0.04,-0.02){12}{\rule{0.5pt}{0.5pt}}
     \end {picture}}

\def\be{\begin{equation}}
\def\ee{\end{equation}}
\def\bea{\begin{eqnarray}}
\def\eea{\end{eqnarray}}

\def\a{\alpha}
\def\b{\beta}

\def\G{\Gamma}

\def\m{\mu}

\title{A Bound on the Entropy of Supergravity?} 
\author{Jan de Boer$^1$, Sheer El-Showk$^1$, Ilies Messamah$^1$ and Dieter Van den Bleeken$^2$

\\

$^1$ Instituut voor Theoretische Fysica, Universiteit Amsterdam, \\
Valckenierstraat 65, 1018XE Amsterdam, The Netherlands\\
\\

$^2$ NHETC and Department of Physics and Astronomy, Rutgers University, \\
Piscataway, NJ 08855, USA \\
}

\abstract{We determine, in two independent ways, the number of BPS
quantum states arising from supergravity degrees of freedom in a
system with fixed total D4D0 charge.  First, we count states
generated by quantizing the spacetime degrees of freedom of
``entropyless'' multicentered solutions consisting of
$\antiDp{0}$-branes bound to a D6$\antiDp{6}$ pair. Second, we
determine the number of free supergravity excitations of the
corresponding AdS$_3$ geometry with the same total charge. We find
that, although these two approaches yield a priori different sets
of states, the leading degeneracies in a large charge expansion
are equal to each other and that, furthermore, the number of such
states is parametrically smaller than that arising from the D4D0
black hole's entropy. This strongly suggests that supergravity
alone is not sufficient to capture all degrees of freedom of large
supersymmetric black holes. Comparing the free supergravity
calculation to that of the D6$\antiDp{6}$D0 system we find that
the bound on the free spectrum imposed by the stringy exclusion
principle (a unitarity bound in the dual CFT) seems to be captured
in the dynamics of the fully interacting but classcial
supergravity equations of motion.}

\begin{document}

\section{Introduction}

The D4D0 BPS black hole was the first four dimensional black hole
to enjoy a microscopic statistical interpretation of its entropy.
In \cite{Maldacena:1997de} it was shown that the five dimensional
uplift of this black hole to a black string is dual to a (0,4)
CFT. If $q_0$ and $p^A$ denote the D0 and D4 charges of the black
hole, then to leading order for large charges the central charge
of the CFT equals $c=d_{ABC}p^A p^B p^C$, with $d_{ABC}$ the
intersection numbers of the Calabi-Yau. We will abbreviate
$d_{ABC} p^A p^B p^C$ by $p^3$ in what follows. If in addition
$-q_0\gg p^3\gg1$, Cardy's formula implies that the CFT has
$e^{2\pi\sqrt{\frac{-q_0 p^3}{6}}}$ BPS states with the same
quantum numbers as the black hole. This coincides exactly with the
gravitational entropy, proportional to the black hole horizon, as
given by the Bekenstein-Hawking formula. Although this
determination of the number of black hole degrees of freedom using
AdS/CFT duality is an amazing and beautiful result it is also
partially unsatisfactory as it leaves a direct understanding of
the black hole states in the strongly coupled or gravity regime
unfulfilled. Motivated by this question we will consider a certain
subset of BPS states that carry the same quantum numbers as the
D4D0 black hole directly in their backreacted supergravity
description.

The BPS states we consider arise from quantizing the supergravity
solutions corresponding to D6$\antiDp{6}$D0 bound
states\footnote{As explained in the bulk of the paper the D6 and
$\antiDp{6}$ carry worldvolume flux leading to a total D4D0 charge
for such a bound state.}, and can be studied using the techniques we developed in
\cite{deBoer:2008zn}. These solutions are horizon-(and hence
entropy-)less\footnote{Note that there is a small degeneracy of
$\chi(CY_3)$ from the degrees of freedom of the D0 in the
compactified dimensions. We neglect this fact in most parts of the
paper, because it only changes the prefactor of the entropy rather
than its scaling in the charges. This is discussed in more detail
in section \ref{freesugra}. }  multicentered configurations
\cite{Denef:2000ar, Bates:2003vx, Bena:2005va, Berglund:2005vb}
that come in discrete sets of continuous families, parameterized
by the possible equilibrium positions of the centers and the
number and charges of the D0 centers. We refer to this continuous
deformation space as a ``solution space'' and in
\cite{deBoer:2008zn} we showed that, due to the presence of
intrinsic angular momentum coming from crossed electric and
magnetic fields, this space is naturally interpreted as (a regular
part of) the phase-space of the system and is thus amenable to
geometric quantization. Using partition function techniques we
calculate the number of states associated to all the possible
solution spaces, classified by partitions of the total D0 charge,
and find that, in the large charge regime, there is an exponential
number, $d_{N,I}$, of D6$\antiDp{6}$D0 BPS states. The precise
result we obtain in this paper depends on the value of the total
D0 charge, $q_0$, relative to the D4 charge cubed $p^3$:
\be
\log d_{N,I}=\begin{cases}\left(\alpha\frac{N^2}{4}\right)^{1/3}\,&\qquad\mbox{if}\
    N\leq I \\
\left(\alpha\,\frac{I}{2}(N-\frac{I}{2})\right)^{1/3}\,&\qquad\mbox{if}\ I\leq N
\end{cases}
\ee
with $I=\frac{p^3}{6}$, $N=-q_0+\frac{p^3}{24}$ and $\a=\frac{3}{4}\zeta(3)$.

This result reveals two interesting facts. First, by comparing to
the number of states of the D4D0 black hole, given above, we see
that for large charges the black hole entropy is exponentially
larger. In this sense the BPS states we considered are extremely
sparse in the set of states forming the black hole. Even though
the set of states we obtain from supergravity is too small to
dominate and thus capture the physics of large black holes, it
does appear to have some rather intriguing physics of its own. As
can be seen from the formula above there appears to be a phase
transition at $N=I$ between two different regimes. It follows from
our calculation that this is essentially due to a restriction on
the orientation and size of the total angular momentum of the
system, acting as a cutoff on the spectrum, leading to a smaller
growth of states once $N>I$. More precisely, by the BPS equations
of motion the total angular momentum is forced to always be
positive, i.e. always pointing from $\antiDp{6}$ to D6 along the
axis through those centers, and to be smaller than $\frac{I}{2}$.

Faced with the result that the BPS states of the D6$\antiDp{6}$D0
supergravity solutions are insufficient to account for the D4D0
black hole entropy, one might wonder if there are other, perhaps
more complicated, zero entropy BPS supergravity solutions that
might generate enough states to account for some finite fraction
of the black hole entropy. The second result of this paper is a
computation indicating that this might not be the case. We compute
the number of multiparticle linearized $\caln=1$ (in 5-d)
supergravity BPS modes around AdS$_3\times$S$^2$, the vacuum of
the (0,4) CFT that contains the black hole ensemble. The result of
this computation is an exponential number of states arising in the
free theory, identical to what we  found for the D6$\antiDp{6}$D0
system discussed above, including the phase transition (though the
latter only emerges after imposing the stringy exclusion principle
- i.e. a unitarity constraint - on the free spectrum).

This might indicate that in the full, interacting theory the D6$\antiDp{6}$D0
states dominate the entropy obtainable from quantizing solutions of 5d $\caln=1$
supergravity. There are, however, many subtleties that could spoil such a hasty
conclusion. First of all, the D6$\antiDp{6}$D0 bound states are solutions of the
fully interacting theory but when more closely analyzed the BPS constraint
equations imply that they are composed primarily of non-interacting bits with only
the D6$\antiDp{6}$ core providing a weak form of interactions between the
centers.  As pointed out in \cite{deBoer:2008fk} these configurations can
essentially be spectrally flowed to a gas of weakly interacting gravitons on an
AdS$_3\times$S$^2$ background.  Thus perhaps it is not so great a surprise that
the degeneracy of such configurations is captured by the free theory.

We can, however, also compute the index in the free theory and one
might expect this to fully agree with the degeneracy of the
interacting theory.  It is not clear, however, that the degrees of
freedom captured by the index in the free theory do not undergo a
phase transition when continued to nonzero coupling so there might
still exist other, more intricate, supergravity solutions that do
have a larger degeneracy than the D6$\antiDp{6}$D0 but that cease
to exist in the free limit. We are not aware of a single solution
that has this property however.

Another interesting observation that follows from the counting in
the free theory is that there the cutoff on the spectrum is
provided by the stringy exclusion principle
\cite{Maldacena:1998bw}, a unitarity bound following from the
(0,4) superconformal algebra, which must be enforced by hand.
Comparing the free calculation and the D6$\antiDp{6}$D0 bound
states we see that the bound given by this stringy exclusion
principle seems to be encoded in the interacting theory through
the BPS equations of motion which put constraints on the positions
of the various centers of the D6$\antiDp{6}$D0 system such that
the total angular momentum is bounded, $J_3\geq 0$. The latter
bound on the angular momentum is equivalent to the stringy
exclusion principle. This is somewhat surprising as it was
previously suggested \cite{Maldacena:1998bw} that supergravity
would not capture this bound. Although we have no explicit proof
that the most general solution of supergravity satisfies this
bound, the fact that it appears to be satisfied by a certain
subset of solutions is certainly intriguing.

Finally we would like to discuss the relevance of our results to
some of the many applications of these or similar multicentered
solutions in the literature. There is a much larger class of
similar solutions that in 5 dimensions correspond to entropy-less
or smooth solutions {\cite{Bena:2005va}\cite{Berglund:2005vb} and
which are asymptotically indistinguishable from black holes/rings.
Such geometries are often referred to (for better or for worse) as
black hole ``microstate'' geometries as they are believed to play a roll in the fuzzball conjecture, see \cite{Mathur:2005zp, Bena:2007kg, Skenderis:2008qn, Mathur:2008nj, Balasubramanian:2008da} for some reviews.   As discussed in
\cite{Bena:2008nh}, even if the vast majority of black hole states
arise from stringy excitations we may still hope to extract useful
information from supergravity states if they are suitably dense in
the black hole Hilbert space.  Precisely what ``suitably dense''
means is not clear but at the very least one would imagine the
entropy of all such configurations should scale in the same way as
the total black hole entropy with the charges (though perhaps with
a different coefficient).  Our results are, at first sight, rather
discouraging for the (supergravity based) microstate program.  This because, at least
in the class of solutions we consider, our results suggest that
generic microstates will not be accessible within supergravity as
these solutions have a parametrically smaller degeneracy than that
required to capture the black hole's entropy. Thus to find a
generic microstate and determine its spatial properties may
require incorporating string or brane degrees of freedom which is
technically quite hard, for an in depth discussion see \cite{Balasubramanian:2008da}.  We pointed out above that there remains
the important disclaimer that we were not able to explicitly show
that there are no other entropy-less supergravity configurations
that might generate a large enough degeneracy, although we have
shown that if such configurations exist they would have to
disappear in the free limit around AdS$_3$.

Another approach to the interpretation of black hole states
\cite{Gaiotto:2004ij,Kim:2005yb} is also related, as shown in
\cite{Denef:2007yt}, to the D6$\antiDp{6}$D0 system studied here.
In \cite{Denef:2007yt,Gimon:2007mha} arguments were given that, if in the
scaling regime the D0-branes of such a D6$\antiDp{6}$D0 bound
state blow up to D2-branes through a version of the Myers effect,
the quantization of these solutions would reproduce the entropy of
the corresponding D4D0 black hole through the same Landau
degeneracies as found in \cite{Gaiotto:2004ij}.  Our results seem
to further confirm that such non-Abelian degrees of freedom are
essential to produce enough degeneracy as our analysis of the
system of \cite{Denef:2007yt} does not incorporate such
non-Abelian stringy effects and yields too few states\footnote{Of
course, non-Abelian degrees of freedom are part of the open string
description of the system, and since our goal is to find a closed
string description of the degrees of freedom we should eventually
avoid non-Abelian degrees of freedom. Such a description could
e.g. be that of fully back-reacted spherical D2-branes, whose explicit form is presently unknown. A first approach to this problem is made in \cite{Levi:xxx}.}. Where in
\cite{Gaiotto:2004ij} each D2 brane with given induced D0 charge
occupied one of $\sim p^3$ landau levels in the Calabi-Yau we find
that in our quantization the addition of a single D0 to the
D6$\antiDp{6}$ systems seems to also yield a growth in the
degeneracy of order $p^3$. The difference in our case is, however,
that the number of states gets suppressed as $1/n!$ for $n$ D0
centers due to the bound on the angular momentum leading to our
observed growth with a power of $1/3$ in the charges, rather than
the power of $1/2$ that the black hole entropy exhibits and which
was reproduced in \cite{Gaiotto:2004ij}. As this is an effect that
emerges only after taking into account the full back reaction of
the D0's and dominates for $n \gg 1$ one might wonder if the
approximation of \cite{Gaiotto:2004ij}, neglecting the
backreaction of the D0's on the geometry, might have to be
corrected in a non-trivial way.

The paper is organized as follows. In section \ref{sec_dipole} we
compute the entropy of asymptotically AdS$_3$ states, first by
restricting to a large class of fully backreacted solutions and
then in the free theory and we show that these match.  We begin by
recalling, in section \ref{deconstruct_setup}, a convenient
coordinatization of the D6$\antiDp{6}$D0 system and quickly review
the quantization of this system following \cite{deBoer:2008zn}. In
section \ref{scaling_partition} we study the partition function of
this system and use it to determine the degeneracy at large
charges.  A similar computation is done for the partition function
of supergravity linearized around a background global
AdS$_3\times$S$^2$ metric in section \ref{freesugra}.  In the free
theory the stringy exclusion principle must be introduced by hand
and changes the growth of states at $N > I$ whereas this
transition emerges naturally from the D6$\antiDp{6}$D0 partition
function.  In section \ref{sec_phys_scaling} we turn out attention
to some other questions of relevance for the physics of these
solutions.  We provide an argument,  complementary to that of
\cite{deBoer:2008zn}, based on AdS/CFT, for the existence of
macroscopic quantum fluctuations in section \ref{sec_macro}.  To
address a long-standing debate regarding whether it is the
components of angular momentum coming from pairs of centers or
only the total angular momentum that is quantized we use some
mathematical results to study the $J=0$ submanifold of a general
phase space in section \ref{sec_j0}. We find that, roughly
speaking, all angular momenta are quantized but that the
respective quantum numbers are not all independent. Some technical
results regarding counting states on our phase spaces regarding as
polytopes of toric manifolds are derived in appendix
\ref{diet-proof}.

\section{Counting dipole halo states}\label{sec_dipole}

In \cite{deBoer:2008zn} the solution space associated with a
D6$\antiDp{6}$ pair (with intersection product $I$ given below)
surrounded by $N$ D0's fixed in the plane orthogonal to the
D6$\antiDp{6}$ axis, a system we will refer to as the dipole halo,
was quantized in the {\em non-scaling} regime and the entropy was
determined to grow as $S \sim N^{2/3}$. The non-scaling regime is
characterized by $N < I/2$ whereas scaling solutions satisfy $N >
I/2$. We remind the reader that scaling solutions are solutions
where the centers, which are labelled by points $x\in\mathbf R^3$,
can approach each other arbitrarily closely (see section~\ref{scso}).

Earlier arguments in the literature \cite{Bena:2006kb, Denef:2007vg, Bena:2007qc, Denef:2007yt} have
suggested that scaling solutions carry vastly more entropy and may
account for a large fraction of the black hole entropy.  Here we
will see that this is not the case, at least for this large class
of solutions.  However, as we will point out, the (leading)
entropy coming from these solutions matches that of free gravitons
in AdS$_3\times$S$^2$.  The change in the leading degeneracy
between the non-scaling and scaling regime seems to precisely take
into account the stringy exclusion principle
\cite{Maldacena:1998bw}, which for a chiral primary in the NS
sector states that $\tilde{L}_0 \leq c/12$.

\subsection{The dipole halo solution space}\label{deconstruct_setup}

In this section we will review some essential formulas and results on the
multicenter solutions of our interest.  We refer readers unfamiliar with the
subject to the earlier literature
\cite{Denef:2000nb,Denef:2002ru,Bates:2003vx,Bena:2005va,Berglund:2005vb,deBoer:2008fk,deBoer:2008zn}.
As was discussed in \cite{Denef:2007yt, deBoer:2008zn} this system manifests two different
behaviors depending on the value of the D0 charge\footnote{Throughout this note
we are going to be sloppy and use D0 instead of $\antiDp{0}$ for brevity.}.

\subsubsection{Charges and constraints}

To be more precise let us first specify what we mean by the charges D6 and
$\antiDp{6}$.  These are D6 branes wrapping the entire Calabi-Yau which carry
lower dimensional brane charge induced entirely from a non-trivial worldvolume
Abelian flux, $F^A= p^A/2$, on the D6 and the opposite flux on the $\antiDp{6}$.
Explicitly, the full charge vector including D6, D4, D2 and D0 charge, in the
same notation as \cite{deBoer:2008zn}, reads:
\begin{itemize}
\item D6, $\G_6=(1,\frac{1}{2}\,p^A,\frac{1}{8}\,D_{A B C} p^B p^C,\frac{1}{48} \,D_{A B C} p^A p^B p^C) = (1,\frac{p}{2},\frac{p^2}{8},\frac{p^3}{48})$.
\item $\antiDp{6}$, $\G_{\bar 6} = (-1,\frac{1}{2}\,p^A,-\frac{1}{8}\, D_{A B C} p^B p^C,\frac{1}{48}\, D_{A B C} {p^A p^B p^C}) = (-1,\frac{p}{2},-\frac{p^2}{8},\frac{p^3}{48})$.
\end{itemize}
where $D_{A B C}$ is the triple intersection of a basis of four
cycles in the Calabi-Yau, and we suppressed some subleading
contributions which depend on the second Chern class of the
Calabi-Yau. In addition to the D6$\antiDp{6}$ pair, the system
consists of an arbitrary number, $n$, of D0's of charge $\GmA =
\{0, 0, 0, -q_a \}$ with all the $q_a$ positive and $\sum_a q_a =
N$, bound to a D6 and $\antiDp{6}$.

To specify the full supergravity solution corresponding to backreacting these
branes it is necessary to fix the moduli at infinity and also specify the
locations of the branes (or centers as we refer to them in supergravity) which
are given by vectors, $\vec{x}_6,\, \vec{x}_{\overline{6}},\, \vec{x}_a \in {\mathbb
R}^3$, in the non-compact spatial directions.  Throughout these
notes we will often find it convenient to work with asymptotically AdS$_3\times$S$^2$
solutions and thus we will work in the decoupling limit of
\cite{deBoer:2008fk} which fixes the asymptotic K\"ahler moduli at $p^A\infty$
(which in this specific case is also a ``threshold point'' \cite{deBoer:2008fk}).

The locations of the centers are not entirely free but must
satisfy so-called ``integrability conditions''
\cite{Bates:2003vx}. It is precisely these conditions that make
the solution space a topologically interesting manifold. In this
case, where we have assumed AdS asymptotics, they take the form
\begin{align}
    -\frac{q_a}{x_{6a}} + \frac{q_a}{x_{\overline{6}a}} &= 0 \label{d6d6d0_c1}\\
    -\frac{I}{x_{6\bar6}} + \sum_a \frac{q_a}{x_{6a}} &= -\beta \label{d6d6d0_c2}
\end{align}
Here $I = - \langle \GmSix, \GmSixb \rangle=\frac{p^3}{6}$ is
given in terms of the total D4-charge $p^A$ of the system and
$\beta = \langle \GmSix, h \rangle$ with $I, \beta > 0$.  Here
$\langle-,-\rangle$ is an anti-symmetric intersection product on
the charge space and $h$ is a vector (in the charge lattice)
specifying the asymptotics of the solution, see e.g.
\cite{Bates:2003vx}.  It is clear from the first line that the
D0's are forced to lie in the plane equidistant from the $D6$ and
$\overline{D6}$, as we are at threshold, and so we can simply
write $x_a := x_{6a} = x_{\overline{6}a}$.

The integrability equations alone are in principle not sufficient
to guarantee the existence of a well-defined solution. For
non-scaling solutions, there are strong arguments
\cite{Denef:2000nb} that existence of the solution is equivalent
to the existence of a so-called attractor flow tree, which does
indeed exist for the \dihalo solutions
\cite{Denef:2007vg,deBoer:2008fk}. For scaling solutions this
argument does not apply, but for the case at hand the main
difference between scaling and non-scaling solutions lies in the
existence of solutions which resemble the D0-D4 black hole very
closely. As the latter have no pathologies, we expect that for
\dihalo systems the integrability conditions are equivalent to the
existence of a solution, but it would be nice to separate check of
this.

\subsubsection{Scaling solutions}  \label{scso}

The two different regimes mentioned above are the non-scaling case
($N< I/2$) which was discussed in detail in section 6 of
\cite{deBoer:2008zn} and the scaling regime ($N\geq I/2$), within
which the centers can approach each other arbitrarily closely. In
figure \ref{totalchargeregions} we show how the scaling and
non-scaling regimes overlap with the so called polar and black
hole regimes. Notice that the scaling solutions we consider do not
cover the region with $I/4<N<I/2$, while black holes exist in this
regime. It is conceivable that other scaling solutions exist, e.g.
with different D6-charges or more than two centers with D6 charge,
which cover this regime, and it would be interesting to explore
this further.

\FIGURE{\includegraphics[scale=1]{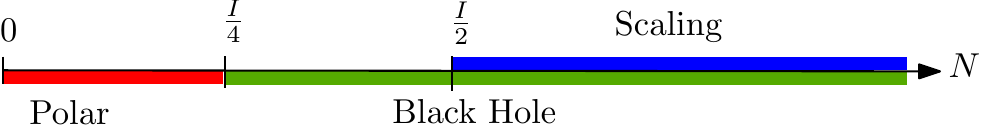}\caption{We
show the different physical regimes of the D6$\antiDp{6}$D0 system
in terms of the total charge carried by the $n$ D0s, $N$. Note
that the total D0-charge is $q_0=\frac{I}{4}-N$ with
$I=\frac{p^3}{6}$. The polar regime corresponds to total charges
for which no single center D4D0 black hole exists while in the
black hole regime there exists a D4D0 black hole with the same
total charges as the D6$\antiDp{6}$D0 system.  Another regime is
what we call the 'scaling' regime in which the D6$\antiDp{6}$D0
solution space contains a point where the different centers become
coincident. }\label{totalchargeregions}}

To see that in the scaling regime the centers can approach each other, let us place all the D0
charge at one center so $q_1=N$ and consider solutions of the form
\be
x_{6\bar6} = \lambda\, I + {\mathcal O}(\lambda^2) \qquad x_{61} = \lambda \,N + {\mathcal
O}(\lambda^2)
\ee
For small $\lambda$ solutions of this form can always be found so long as $N
\geq I/2$; the latter requirement coming from the fact that $x_{6\bar6}$ and
$x_{61}$ are coordinate separations and must satisfy triangle inequalities.  As
$\lambda \rightarrow 0$ the {\em coordinate} distance between the centers goes
to zero and the centers coincide in coordinate space.  In physical space,
however, warp factors in the metric blow up generating a deep throat that keeps
the centers a fixed metric distance apart even as $\lambda \rightarrow 0$.
Outside of this arbitrarily deep throat the solutions is almost
indistinguishable from a D4D0 black hole.  This regime is thus of great physical
relevance and e.g. in \cite{Denef:2007yt} it was conjectured to correspond to the deconstruction of a D4D0
black hole.

\subsubsection{Symplectic form}
We are now ready to review the construction of the associated symplectic form.
Recall \cite{deBoer:2008zn} that once the symplectic form on the solution space
(parameterized by the locations of centers satisfying
(\ref{d6d6d0_c1})-(\ref{d6d6d0_c2})) has been found it can be used to quantize the
system using methods of geometric quantization.  To make this paper somewhat
self-contained we begin by reproducing the relevant part of section 6 of
\cite{deBoer:2008zn}. For a derivation and more background the reader is
referred to \cite{deBoer:2008zn}.

\FIGURE{\includegraphics[scale=1.8]{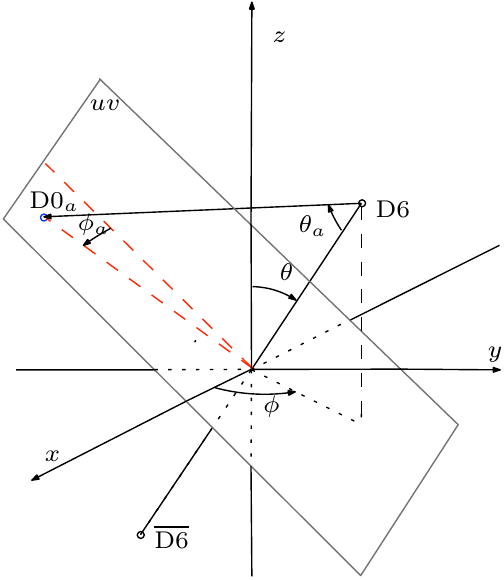}
\caption{The coordinate system used to derive the D6$\antiDp{6}$-N D0
symplectic form.  The coordinates ($\theta$, $\phi$) define the orientation of
the ($\hat{u}, \hat{v}, \hat{w}$) axis with respect to the fixed, reference,
($\hat{x}, \hat{y}, \hat{z}$) axis.  The D6$\antiDp{6}$ lie along the $\hat{w}$
axis (with the origin between them) and the D0's lie on the $\hat{u}$-$\hat{v}$
plane at an angle $\phi_a$ from the $\hat{u}$-axis.  The radial position of
each D0 in the $\hat{u}$-$\hat{v}$ plane is encoded in the angle $\theta_a$
(between $\vec{x}_{6\overline{6}}$ and $\vec{x}_{6a}$).  }\label{fig_coords}}

An explicit expression for the symplectic form  can be obtained
using the following coordinate system \cite{deBoer:2008zn}.  We
define an orthonormal frame $(\hat{u}, \hat{v}, \hat{w})$ fixed to
the D6$\antiDp{6}$ pair, such that the D6$\antiDp{6}$ lie along
the $w$ axis and with the D0's lying in the $u$-$v$ plane.
Rotations of the system can then be interpreted as rotations of
the $(\hat{u}, \hat{v}, \hat{w})$ frame with respect to a fixed
$(\hat x,\hat y,\hat z)$ frame. We will parameterize the choice of
$\hat{w}$-axis in the standard fashion by two angles,
$(\theta,\phi)$.  We can furthermore specify the location of the
$a$'th D0 with respect to D6$\antiDp{6}$ pair by two additional
angles, $(\theta_a,\phi_a)$. The first angle, $\theta_a$, is the
one between $\vec{x}_{6\overline{6}}$ and $\vec{x}_{6a}$, while
$\phi_a$ is a polar angle in the $u$-$v$ plane .  Our $2n+2$
independent coordinates on solution space are thus $\{\theta,
\phi, \theta_1, \phi_1, \dots, \theta_n, \phi_n\}$.  This
coordinate system is depicted in figure \ref{fig_coords}.

The standard Euclidean coordinates of the centers are then given in terms of the angular coordinates by
\begin{align}
    \vec{x}_6 &= \frac{j}{\beta} \hat{w} \qquad &\hat{u} &= \cos \phi\, \hat{x} -
    \sin \phi \,\hat{y} \label{dipole_c1} \\
    \vec{x}_{\overline{6}} &= - \frac{j}{\beta} \hat{w} \qquad &\hat{v} &= \sin
    \phi \cos \theta \,\hat{x} + \cos \phi \cos \theta \,\hat{y} - \sin\theta \,
    \hat{z} \label{dipole_c2} \\
    \vec{x}_a &= \frac{j \, \tan \theta_a}{\beta} (\cos \phi_a \hat{u} + \sin
    \phi_a \hat{v}) \qquad &\hat{w} &= \sin\phi \sin\theta \, \hat{x} + \cos\phi
    \sin\theta \, \hat{y} + \cos \theta \, \hat{z} \label{dipole_c3}
\end{align}
The angular momentum, $j(\theta_a)$, is a function of the other
coordinates rather than an independent coordinate, and is given by
\be \label{deconst-ang-moment}
j = \frac{I}{2} - \sum_a q_a \cos\theta_a\,.
\ee
Using this explicit coordinatization the symplectic form turns out to be \cite{deBoer:2008zn}:
\be \label{deconst-symp-form}
\Omega = -\frac{1}{4} d \biggl[ 2j \cos \theta \, d\phi + 2 \sum_a
q_a \cos\theta_a \, d\phi_a\biggr]
\ee
with $d$ denoting the exterior derivative.

The symplectic form (\ref{deconst-symp-form}) is non-degenerate on
the BPS solution space parameterized by the locations of the
centers implying that the latter is in fact a phase space.  By
virtue of arguments in \cite{deBoer:2008zn} this space can be
quantized in its own right, ignoring the much larger non-BPS
solution space in which it is embedded, and from this treatment
one might hope to extract information about the BPS states of the
full theory (including at least the number of such states).

Note that, as is manifest from our angular coordinatization, the phase space is
actually toric with a U(1)$^{n+1}$ action coming from $\phi$ and the $n$
$\phi_a$'s.  This is a consequence of the fact that the D0's are mutually
non-interacting; their sole interaction is via the D6$\antiDp{6}$.  This toric
structure greatly simplifies the quantization of the solution and was essential
in \cite{deBoer:2008zn}.

\subsubsection{Physical picture}\label{sec_phys_pic}

As much of the subsequent presentation will be a rather technical treatment
of the phase space we would like to lend the reader some intuition.  We begin by
recalling \cite{Bates:2003vx} that the angular momentum carried by these
solutions is
\be
\vec{J} = \sum_{i<j} \vec{J}_{ij}\,, \qquad
\vec{J}_{ij}:= \frac{\langle \Gamma_i, \Gamma_j\rangle \vec{x}_{ij}}{2 \, r_{ij}}
\ee
where now $i, j$ run over all centers, including the D6s.  Thus each pair of
centers contributes angular momentum $\vec{J}_{ij}$ to the total.  The length of
these vectors is fixed to $\langle \G_i, \G_j \rangle/2$ but their direction is
not fixed.  The dependence on the intersction product $\langle \G_i, \G_j
\rangle$, pairing electric and magenetic sources, reflects the fact that
this angular momentum is carried by the electromagnetic field and is due to
crossed electric and magnetic fields.  Since the D0's have vanishing intersection
product with each other there are only $(2n+1)$ momenta vectors:
$\vec{J}_{6\overline{6}}$, $\vec{J}_{6a}$, and $\vec{J}_{\overline{6}a}$.

As we will see, our quantization can essentially be understood as
quantizing the direction of these vectors, or more precisely the
size of their projection on a given ``z-axis'', yielding familiar
angular momentum multiplets.  Naively the phase space of these
angular momentum vectors is the direct product of $(2n+1)$
two-spheres and the number of states is just the product of the
factors $(2|\vec{J}_{ij}|+1)$ from each multiplet. The geometric
origin of the momenta (i.e. endpoint of multiple vectors fixed to
be the same center), however, as well as the constraint equations
(\ref{d6d6d0_c1})-(\ref{d6d6d0_c2}) fix the possible relative
orientations of the different angular momentum vectors. As a
result not all states of the full free angular momentum multiplets
are allowed. Rather, the correct phase space is now a more
complicated fibration of spheres of varying size and, although
intuitively it is still insightful to think of the states as part
of ``angular momentum multiplets'', they now only fill out a
constrained subspace of the product of the full multiplets.  For
instance, since $\vec{J}_{6a}$ and $\vec{J}_{\overline{6}a}$
always end at the same point, their orientation relative to the
$w$-axis is not independent so, rather than two angular momentum
multiplets, these vectors yield only a single multiplet (the
diagonal multiplet in their free product).

The best way to get some intuition for this is to consider the symplectic
form on the phase space of our system.  Using the coordinate system
(\ref{dipole_c3}) and introducing the notation

\begin{align}
    \vec{J}&=J_z \hat{z}+{J}_y \hat{y}+{J}_x \hat{x}\\
    \vec{J}_{6a}&=\vec{J}^a={J}^a_w \hat{w}+{J}^a_v \hat{v}+{J}^a_u \hat{u} \\
    \vec{J}_{\overline{6}a}&=\vec{J}^\ova={J}^\ova_w \hat{w}+{J}^\ova_v
    \hat{v}+{J}^\ova_u \hat{u}
\end{align}
we can cast the dipole halo symplectic form in a more suggestive form
\begin{align}
\Omega &= -\frac{1}{2}  \biggl[ d J_z  \wedge d\phi +
\sum_a  d J^a_w\,   \wedge d\sigma_a +
\sum_a  d J^\ova_w\,   \wedge d\sigma_\ova\biggr]\,.
\end{align}
with $|\vec{J}^a| = |\vec{J}^\ova| = q_a/2$.  Because
$\vec{J}^a$ and  $\vec{J}^\ova$ are related by the location of the D0 they end
on the last two terms above can be combined yielding
\begin{align}
\Omega &= -\frac{1}{2}  \biggl[ d J_z  \wedge d\phi +
2 \sum_a  d J^a_w\,   \wedge d\sigma_a\biggr]\,.
\end{align}
If there were no other constraints, the $J^a_w$ would independently be able to take
values between $\pm|\vec{J}^a|$.  However, as we will discuss below, they have
to satisfy the bounds $J^a_w>0$ and $2 \sum_a J^a_w\leq I/2$, leading to a more
intricate phase space with a Hilbert space that is no longer a product of
``free'' angular momentum multiplets. There is also another angular momentum
multiplet, coming from the total angular momentum $\vec{J}$,
and this gives rise to a full multiplet with $-|J|<J_z<|J|$.  The size
of $\vec{J}$, however, depends on the $J^a$ (even classically).  Thus, each state in
the Hilbert state labelled by $J^a$ quantum numbers will be tensored with a $J$
multiplet corresponding to the total $\vec{J}$ associated to its $J^a$ quantum
numbers (via $J = I/2 - 2 \sum_a J^a_w$).

This is the intuitive physical picture for which we develop a precise
mathematical treatment in the next subsection, using the observation of
\cite{deBoer:2008zn} that the phase space is a toric manifold.  The upshot is,
however, that we are doing nothing more than quantizing angular momentum
variables, but ones that are non-trivially connected and constrained.

There is a second physical phenomena that only appears when quantizing the
system, which we would like to highlight here.  As stressed above, we are in
essence quantizing the classical angular momentum of the system. However, when
we quantize we need to take into account the intrinsic spin of the particles
involved, as was beautifully explained in \cite{Denef:2002ru}. As pointed out
there, the centers are superparticles containing excitations in various spin
states. Due to the presence of magnetic fields, however, the lowest energy BPS
state is a spin half state, where energy is gained by aligning the intrinsic
magnetic dipole moment with the magnetic field. The situation is sketched for
our dipole halo system in figure \ref{spins}. Including these quantumcorrections the size of the total angular momentum is given by
\be
J=\frac{I-1}{2}-\sum_{a}\left(q_a\cos\theta_a+\frac{1}{2}\right)\,.
\ee

\FIGURE{\includegraphics[scale=1]{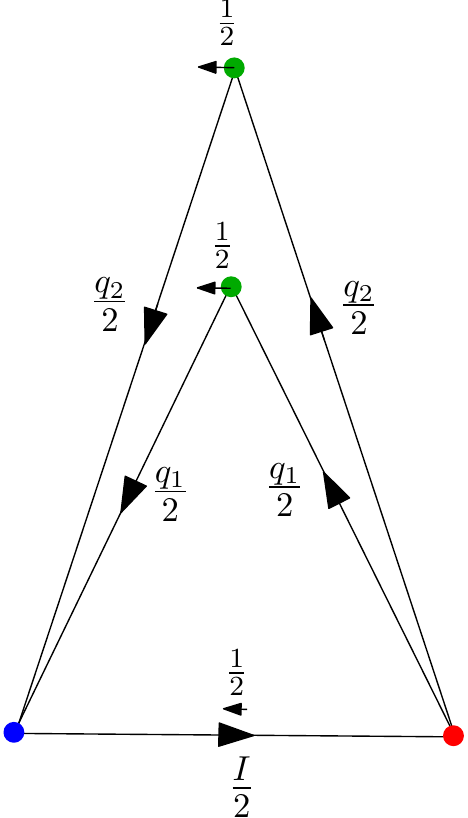}
\caption{In this figure all contributions to the total angular momentum are
shown. The large arrows denote the classical angular momenta carried by the
electromagnetic field. They are proportional to the intersection products of the
charges, D6 (red), $\antiDp{6}$ (blue) and D0 (green). The small
arrows extending from the D0's represent their spin aligning with the dipole
magnetic field sourced by the D6$\antiDp{6}$-pair, the small arrow in the bottom
is the spin of the center of mass multiplet of the D6$\antiDp{6}$-pair aligning
with the magnetic fields.}\label{spins}}

These spins are especially important when considering classical scaling
solutions, as we discuss further in section \ref{sec-morethree}.

\subsection{The dipole halo states}\label{scaling_partition}

As mentioned earlier the system under study manifests two regimes.  The
non-scaling one ($N<I/2$) was studied in detail in \cite{deBoer:2008zn}, and
here we will extend that calculation so that the whole allowed range, $N>0$, is covered.

In this section we will count states using techniques of geometric quantization
of the supergravity solution spaces developed in \cite{deBoer:2008zn}; in the
next section we will compare this to the calculation of free supergravity states
on AdS$_3$ and see that the two results agree beautifully.

\subsubsection{States and polytopes}

In our previous work \cite{deBoer:2008zn} we showed that for the \dihalo system
the solution space is a toric manifold.  This allowed us to construct all the
normalizable quantum states explicitly. In this paper we will be less interested
in the explicit form of the wave-functions than in their number. In
appendix \ref{diet-proof} we show how the number of states can be easily
obtained from the combinatorics of the toric polytope. We will not review the
technology of geometric quantization of toric manifolds here but instead refer
the reader to Appendix B of \cite{deBoer:2008zn} and Appendix \ref{diet-proof}
of this note.

From the symplectic form (\ref{deconst-symp-form}) we read off the coordinates
on the polytope
\begin{equation}\label{ycoords}
   y = j \; \cos \theta \,, \qquad \qquad y_a = q_a \, \cos\theta_a \geq 0
\end{equation}
So we see that the polytope is bounded by the inequalities
\be
-j\leq y\leq j\,,\qquad 0\leq y_a\leq q_a\label{standcond}
\ee
and furthermore the requirement that the angular momentum is positive
\be
  j=\frac{I}{2}-\sum_{a} y_a \geq 0\,.\label{scalingcond}
\ee
It is this last condition that differentiates the non-scaling
regime $N=\sum_a q_a<I/2$ from the scaling regime $N \geq I/2$. In
the former range the condition (\ref{scalingcond}) is redundant in
the definition of the polytope as it is automatically satisfied
for all values of $x_a$ allowed by the other constraints
(\ref{standcond}). In case $N>I/2$ the constraint
(\ref{scalingcond}) actually becomes essential and can make some
of the constraints (\ref{standcond}) redundant, although this
depends on the specific values of the $q_a$. What is shared by all
the solution spaces in the scaling case is that it is possible to
approach the point where all centers coincide arbitrarily closely,
which automatically implies that $j$ has to approach zero. When
this happens, an infinitely deep scaling throat forms in
space-time \cite{Denef:2002ru, Bena:2006kb}. For more than a
single D0 center there are however different types of solution
spaces with a scaling point, depending on the specific values of
the charges $q_a$. We show all the different possible polytope
topologies for the case with two D0 centers in figure
\ref{4cpolytopes}, clearly the number of topologies grows very
fast with the number of D0-centers.

\FIGURE{\includegraphics[scale=1.1]{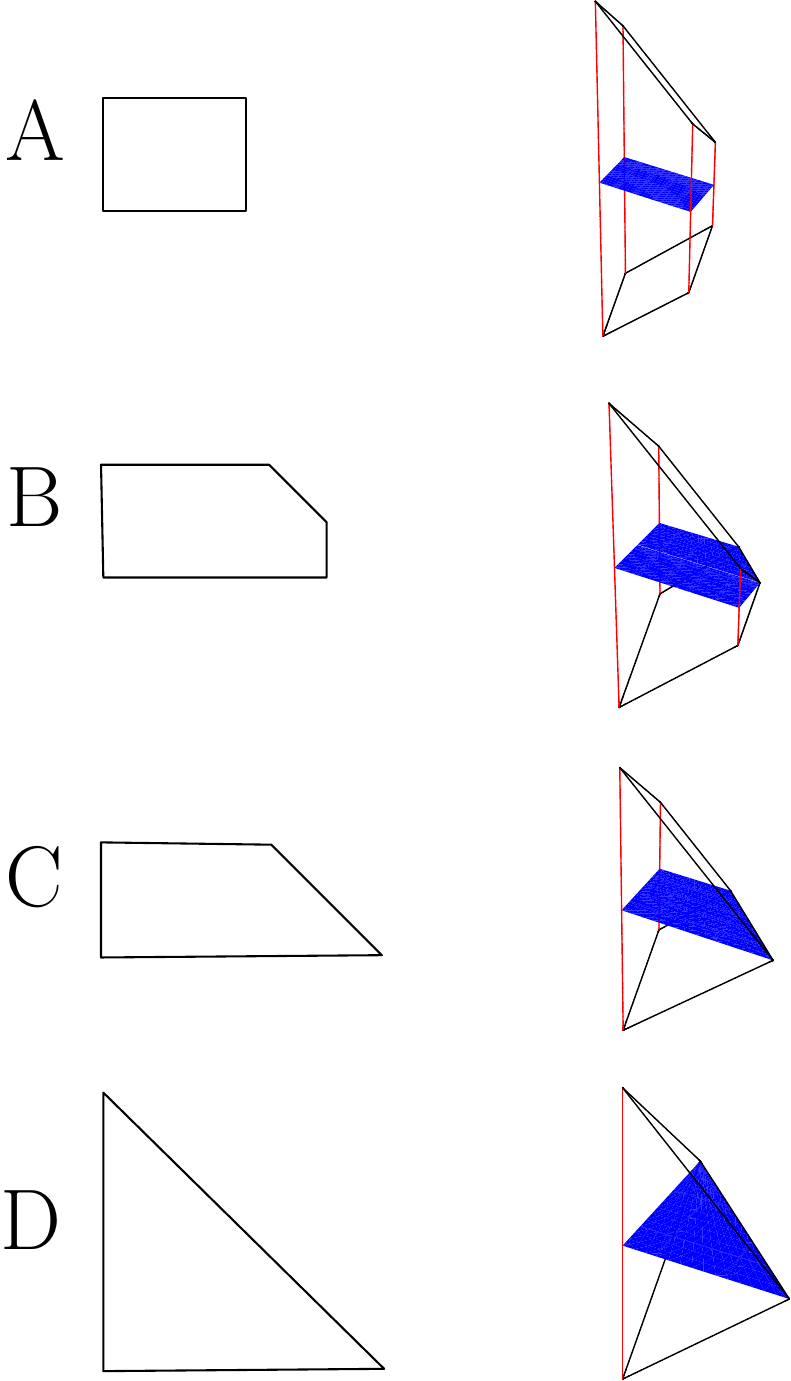}
\caption{These are the different types of polytopes corresponding to a
D6$\antiDp{6}$D0 system with 4 centers. On the left the 'base' polytope
determined by the coordinates $y_1$ and $y_2$ with $y=0$ is shown, on the right also the
fiber spanned by the coordinate $y$ is included, the edges along this direction
are drawn in red, the surface $y=0$ is shown in blue. See (\protect\ref{ycoords}) for a definition of the
coordinates. The different cases correspond respectively to: Case A (non-scaling) $q_1+q_2<\frac{I}{2}$, Case B (scaling) $q_1+q_2\geq\frac{I}{2}$ and $q_1,\,q_2\leq\frac{I}{2}$, Case C (scaling) $q_2\leq\frac{I}{2}\leq q_1$, and Case D (scaling) $\frac{I}{2}\leq q_1,\,q_2$. So we see that from 4 centers onward there are different types of scaling
polytopes, a feature that was absent for three scaling centers.
}\label{4cpolytopes}}

Given the defining inequalities (\ref{standcond}) and (\ref{scalingcond}) we can
use eqn. (\ref{states=volume}) from the appendix (see also the example
containing eqn.  (\ref{defscalepoly})) to see that there is a unique quantum
state corresponding to each set of integers $(m,m^a)$ satisfying
\begin{eqnarray}
0\leq m^a\leq q_a - 1\,,\qquad \sum_a (m^a+\frac{1}{2})\leq \frac{I-1}{2}\,,\\
- \left[ \frac{I - 1}{2} - \sum_{a} \left( m^a + \frac{1}{2} \right) \right] \leq m + \frac{1}{2} \leq \left[ \frac{I - 1}{2} - \sum_{a} \left( m^a + \frac{1}{2} \right) \right] \label{nonorbcond}
\end{eqnarray}
The $(m, m_a)$ above are simply quantized angular momenta corresponding to
quantizing the angles $(\theta, \theta_a)$ appearing in
(\ref{standcond})-(\ref{scalingcond}).  The half-integral shifts are related to
the fermionic nature of the centers as discussed in \cite{Denef:2002ru, deBoer:2008zn} and the coupling to the extrinsic spin, as explained at the end of section \ref{sec_phys_pic}.

To be precise the constraints above only hold under the assumption that all D0
centers carry different charges, $q_a$. To relax this assumption we introduce
integer multiplicities, $n_a$, for each charge $q_a$ so that $N=\sum_a n_a q_a$
and $n=\sum_a n_a$.  We now have to take into account the quantum
indistinguishability of these (fermionic) particles. As explained in detail in
\cite{deBoer:2008zn} this translates to taking the appropriate orbifold of the
polytope or, in terms of (\ref{nonorbcond}), augmenting the $m^a$ by an
additional label $i_a$ running from $1,\ldots, n_a$ and requiring them to
satisfy
\begin{eqnarray}
  0 \leq m^a_1 < m^a_2 < .... < m^a_{n_a}<q_a\, \,, \qquad \qquad
  \sum_{a,i} \left(m^a_{i_a} + \frac{1}{2}\right)  \leq \frac{I - 1}{2} \label{bh-deconst-cond-3} \\
          - \left[ \frac{I - 1}{2} - \sum_{a,{i}} \left( m^a_{i_a} + \frac{1}{2}
          \right) \right] \leq m + \frac{1}{2} \leq \left[ \frac{I - 1}{2} -
          \sum_{a,i} \left( m^a_{i_a} + \frac{1}{2} \right) \right] \label{bh-deconst-cond-4}
\end{eqnarray}
These constraints are fermionic, enforcing Pauli exclusion of indistinguishable
centers.  Note also that they reduce to (\ref{nonorbcond}) if all the $n_a =
1$.

\subsubsection{The D6$\antiDp{6}$D0 partition function}

In this section we will count the combined number of supergravity states $d_N$
of all \dihalo systems with total charge $(p^A,\frac{p^3}{24}-N)$. More precisely
we will calculate the leading term of $S(N)=\log d_N$ in a large $N$ expansion.
We will notice there are two phases depending on the relative value of $I=p^3/6$
and $N$, separated by a transition at $N=I$. The first phase was already found
in \cite{deBoer:2008zn} where the counting was performed in the regime $N<I/2$.
For larger $N$ the appearance of scaling solutions slightly complicates the
counting but can still be performed as shown below. What is interesting is that
the existence of scaling solutions seems only to become dominant at $N=I$ where
a phase transition occurs.

As we previously pointed out, in the scaling regime there is an additional
constraint that complicates the polytope and makes the counting of integer
points inside slightly more difficult. We will find it convenient not to calculate
a fully explicit generating function $Z$, as we did before in \cite{deBoer:2008zn}. Instead,
since we are only interested in the large $N$ regime, it will be sufficient for
us to find the leading term of $\log Z$ in a large $N$ expansion.

The complication in the scaling regime arises because of the second constraint
in equation (\ref{bh-deconst-cond-3}). To proceed let us introduce the quantity
  \be
  M = \sum_{a,i} \Big(m^a_{i_a}  + \frac{1}{2}\Big)\,,\label{Mdef}
  \ee
As the $m_{i_a}^a$ are the discrete analogues of the classical $q_a\cos\theta_a$
the interpretation of $M$ is as the amount of angular momentum carried by the D0
centers (which, by the integrability constraints
(\ref{d6d6d0_c1},\ref{d6d6d0_c2}), is always opposite in direction to the
angular momentum carried by the D6$\antiDp{6}$ pair):
\be
M=\frac{I}{2}-\frac{1}{2}-J\,.\label{meaning}
\ee
Both the $\frac{1}{2}$ in the above formula and in (\ref{Mdef})
arise due to the spin contributions to the quantum mechanical
angular momentum (see the end of section \ref{sec_phys_pic} for a
detailed explanation).

Now if we succeed in calculating the degeneracy $d_{N,M}$ as a function of
$M$, then the full degeneracy will be
  \be \label{ang-cond}
     d_{N} = \sum_{M=1/2}^{(I-1)/2} d_{N,M}
  \ee
The full degeneracy will clearly be less than $I/2$
times $d_{N,M'} $ where $M'$ is the value of $M$ which maximizes $d_{N, M}$.
Thus instead of calculating the sum it will be sufficient for us to find the
$M'$ that maximizes $d_{N, M}$ because
\be
S(N)=S(N,M')+\Delta S \,,
\ee
where we have defined
\be
\Delta S=\log\sum_{M=1/2}^{(I-1)/2}e^{S(N,M)}S(N,M')\leq\log I\, ,
\ee
so, as long as the leading entropy is a power-law (rather than a logarithm) in the
charges, we can find the leading term in $S(N)$ by calculating $S(N,M)$ and
maximizing over $M$.

As we will now show it is not too hard to calculate a generating function for $d_{N,M}$

\be
   \calz (q,y) = \sum_{N,M}^{\infty} d_{N,M} \, q^N\,y^M\,.\label{partf}
\ee
Note that this does not reduce to a generating function for $d_N$ by setting
$y=1$ as in this generating function we sum over $M=1,\ldots,\infty$ while in the
case of interest the range of $M$ is restricted.

Let us derive an expression for (\ref{partf}) by approximating it in a few steps.
A first key ingredient is that for a partition of $N = \sum_a n_a q_a$, one has
\be \label{part-1}
   0\leq m_1^a <...<m^a_{n_a}<q_a\,, \qquad M = \sum_{a,i} \Big(m_i^a + \frac{1}{2}\Big)
\ee
Forgetting for the moment about $N$, the above relation is just a fermionic partition of $M$. This is given by
\be\label{fermion-part}
  \calz_{ferm} = \prod_{l\geq1} \left(1 + y^{l-1/2}\right)
\ee
We need to reintroduce the information about $N$. To do so
remember that the sole role of the partition of $N$ is to specify
the number of $m_i^a$ above. Keeping this key point in mind we
proceed in two steps. First assume that we have $n$ centers with
the same charge $k$ only ($N = n k$), then it is easy to see that
the appropriate modification of $\calz_{ferm}$
(\ref{fermion-part}) is
\be\label{interm-part}
  \calz_{int} = \prod_{1\leq l\leq k} \left(1 + q^k y^{l-1/2}\right)
\ee
this comes about because in expanding the expression above the number of centers
in each term is simply the number of $q^k$ that appear in it. The product over
possible $l$ is then a reflection of the constraint (\ref{part-1}). Now to
generalize to an arbitrary partition of $N$ we take a product of the above
expression over all possible $k\geq1$. This yields the \emph{core} generating function
\be\label{core-part}
  \calz_0 = \prod_{ k\geq 1,1\leq l\leq k} \left(1+q^k y^{l-1/2}\right)
\ee
To get the actual generating function we include the contribution from $m$ in
equation (\ref{bh-deconst-cond-4}). The generating function is then
\be \label{bh-deconst-gen-funct}
\calz = (I - 2 y\partial_y) \, \calz_0 = (I - 2 y\partial_y) \prod_{k\geq 1,1\leq l\leq k} \left(1+q^k y^{l-1/2}\right)
\ee
In evaluating the leading contribution to the entropy we can neglect the overall
multiplicative factor because it will be subleading.  Thus we focus on
$\calz_0$.

\subsubsection{The entropy and a phase transition}\label{sec_D6D6D0_entropy}

As is familiar from thermodynamics we can study the large energy regime by
evaluating the partition function at large temperature. We introduce the potentials $\b$ and $\m$ through
   $$ q = e^{-\beta}\,, \qquad y= e^{-\mu} $$
and can then look for the behavior of the entropy for $\beta,\mu \ll 1$.
\bea\label{freeenergy}
  \log \calz_0 &=& \sum_{k\geq 1,1\leq l\leq k} \log \left(1+q^k y^{l-1/2}\right) \nonumber\\
               &=& \sum_{n\geq 1} \left(\frac{(-1)^{n+1}}{n} \left[\sum_{k\geq 1} q^{n k} \left(\sum_{l=1}^{k} y^{n(l-1/2)}\right)\right] \right)\nonumber\\
               &=& \sum_{n\geq 1} \left(\frac{(-1)^{n+1}}{n}\, \frac{y^{n/2}}{1-y^n}\, \left[\sum_{k\geq 1} q^{n k}\, (1 - y^{n k})\right] \right)\nonumber\\
               &=& \sum_{n\geq 1} \left(\frac{(-1)^{n+1}}{n}\, \frac{q^n\,y^{n/2}}{(1-q^n)\, (1 - q^n \, y^n)} \right) \nonumber\\
               &\sim& \left(\sum_{n>1} \frac{(-1)^{n+1}}{n^3} \right)\, \frac{1}{\beta\,(\mu+ \beta)} =:\frac{\alpha}{\beta\,(\mu+\beta)}
\eea
with $\a=\frac{3}{4}\zeta(3)$.
Using the above relation we find
\bea
    N &=& - \partial_{\beta} \log \calz_0 \sim
	\frac{\alpha(\mu+2\beta)}{\beta^2\,(\mu+ \beta)^2} \label{Nexp}\\
    M &=& - \partial_{\mu} \log \calz_0 \sim \frac{\alpha}{\beta\,(\mu+
	\beta)^2} \label{Mexp}
\eea
From the equations above it follows that the approximation is valid for $N,M
\gg1$, which is exactly the regime we are interested in. Furthermore the
relative size between $M$ and $N$ is determined by the ratio $\mu/\beta$ as
\be
N/M=2+\frac{\mu}{\beta}\,.
\ee
The entropy in the large $M,N$ regime then reads
\be
  S(N,M) = -\log \calz_0 + \beta N + \mu M \sim \frac{\alpha}{\beta\,(\mu+ \beta)} \sim \left(\alpha\,M\,[N-M]\right)^{1/3}\,. \label{legendre}
\ee
Maximizing $S(N,M)$ over $M$ in the range\footnote{Note that we are interested
in the large charge limit $I\gg1$, so throughout the paper we will often neglect
quantum mechanical shifts of $1/2$ to $I$} $1/2<M<I/2$ we find that
\be
S(N)=\begin{cases}\left(\alpha\frac{N^2}{4}\right)^{1/3}\,&\qquad\mbox{if}\
    N\leq I \label{entropy2}\\
\left(\alpha\,\frac{I}{2}(N-\frac{I}{2})\right)^{1/3}\,&\qquad\mbox{if}\ I\leq N
\end{cases}
\ee
The most entropic configuration always has $M' = N/2$ until $N = I$ and then the
bound (\ref{scalingcond}) restricts $M'=I/2$.  Thus most entropy is realized by
low angular momentum states (remember $J \sim I/2-M$) and, deep in the scaling
regime where $N>I$, most of the entropy is given by the $j=0$ states.

The saddle point approximation used to obtain eqn. (\ref{entropy2}) is only
valid for charges $N \lesssim I^2$ because the discussion above shows we are interested in $M\approx\frac{I}{2}$ and in that regime $N \gtrsim I^2$ is not
consistent with $\mu, \beta \ll 1$, as can be seen from
(\ref{Nexp})-(\ref{Mexp}).  We will presently focus on the regime $N \gg I$
which is still consistent so long as their ratio does not scale with $I$ in the large charge regime one considers.

For $N \gg I$ Cardy's formula implies the leading entropy of the associated
black hole grows as \cite{Maldacena:1997de}
\be
S_{BH}(N, I) \sim 4\pi \sqrt{\frac{N\, I}{4}}
\ee
Thus the D6$\antiDp{6}$D0 configurations we are considering do not exhibit the
correct growth of entropy as a function of the charges to dominate the black
hole ensemble, especially for large charges they are parametrically subleading.

Associated with the change from the first to the second line of
(\ref{entropy2}) appears to be a phase transition
occurring at $N=I$.  In this phase transition we seem to move from
an asymmetric phase, $\langle j \rangle \neq 0$, to a symmetric
phase $\langle j \rangle = 0$, note that this transition is not
$C^\infty$ but still continuous.  It is not immediately clear that
any physical meaning should be ascribed to this ``phase
transition'' since these configurations are not the dominant
constituents of this sector of the BPS Hilbert space.  Curiously,
however, this seems to mirror the phase transition of
\cite{deBoer:2008fk}.  Although the latter was analyzed for
different constituents centers, if we simply equate the total
charges of the two systems then the critical point of
\cite{deBoer:2008fk} would be at $N \approx I/4$ and would
correspond to a transition from a phase with $\langle j \rangle
\neq 0$ to a $\langle j \rangle = 0$ phase as $N$ increases (note
that here there is a discontinuous jump in $\langle j\rangle$). It
is both curious and interesting that the set of states we
obtained, while relatively sparse in the overall Hilbert space,
nonetheless exhibits a non-trivial phase structure that even seems
to qualitatively share some of the structure of the full theory.

In the regime $N \gg I$ of \cite{Maldacena:1997de}, the number of
states we obtained was substantially smaller than total number of
BPS states of the conformal field theory. One may therefore wonder
whether other solutions of supergravity exist with the same
asymptotic charges and which could account for the missing states,
or whether this is the best supergravity can do. Such additional
solutions could look like complicated multi-centered solutions of
the type we have been considering, or be of an entirely different
form. To address this question we will now compute the spectrum
and degeneracy of a gas of free supergravitons in
AdS$^3\times$S$^2$. As we will argue, this will provide an
estimate for the maximal number of states we might expect to be
obtainable from supergravity. It turns out that this computation
yields a result whose asymptotic expansion agrees precisely with
the number of \dihalo states, which supports the claim that the
supergravity does not give rise to significantly more states in
addition to those that we described.

\subsection{Free supergravity estimate}\label{freesugra}

In the previous section we calculated the number of BPS states in a given
D4D0-charge sector that can be associated to configurational degrees of freedom
of a \dihalo system of that same total charge. As we pointed out, there is an
exponential number of states leading to a macroscopic statistical entropy.
However the entropy scales with a different power of the charges than the D4D0
black hole entropy, making it parametrically subleading in the large charge
supergravity limit. In other words, although we found very many \dihalo states
the corresponding single center black hole still has exponentially more of them,
indicating that these are not generic states of the black hole.

One might still wonder, however, if this is due to our restriction to a specific
set of smooth multicenter solutions and if perhaps a larger number of states can
be found by quantizing more complicated multicentered configurations. In this
section we will give some non-trivial evidence that this is \textbf{not} the
case and that the black hole degrees of freedom have to be sought outside of
supergravity. An example of such states could be those of the proposal
\cite{Gaiotto:2004ij,Raeymaekers:2007ga,Denef:2007yt,Gimon:2007mha} or the
possibly related setup of \cite{Bena:2008nh,Bena:2008dw}. Roughly speaking the
degrees of freedom in these pictures seem to reside in non-abelian D-brane
degrees of freedom; see also \cite{Balasubramanian:2006gi}.

The approach we take to get a ``bound'' on the degrees of freedom
coming from supergravity states is to exploit the fact that both
the D4D0 black hole and the \dihalo system (and its
generalizations) can be studied in asymptotically AdS space via
the decoupling limit of \cite{deBoer:2008fk}. In this context, the
counting of the previous section corresponds to counting
backreacted supergravity solutions with the same asymptotics as
the D4D0 BTZ black hole, whereas here we will simply count free
supergravity modes in empty AdS.  The advantage of working in this
limit, where the supergravity fields become free excitations
around a fixed AdS$_3\times$S$^2\times$CY$_3$ background, is that
it becomes relatively easy to count them. Free supergravitons
organize themselves in representations of the (0,4) superconformal
isometry algebra, and we merely need to determine the quantum
numbers of the highest weights of the representations. This can be
done following e.g. \cite{Larsen:1998xm,deBoer:1998ip} by
performing a KK-reduction of
eleven dimensional supergravity fields on the compact
S$^2\times$CY$_3$ space\footnote{Note that we will assume the size
of the CY$_3$ to be much smaller than that of the S$^2$ so that we
will only consider the massless spectrum on the CY, while keeping
track of the full tower of massive harmonic modes on the sphere.}
to fields living on AdS$_3$.  The supergravity spectrum can then
be determined using pure representation theoretic methods, in
terms of the massless field content of the KK reduction of
M-theory on the Calabi-Yau manifold.

\subsubsection{Superconformal quantum numbers}\label{sqn}

We want to compare the number of states we found by counting the
possible configurational degrees of freedom of a \dihalo system to
the number of chiral primaries given by KK reduction of 5d
supergravity in the free field limit.  To make this comparison as
clear as possible let us first translate the conserved four
dimensional charges of the solutions, as presented in the previous
section, to quantum numbers under the (0,4) superconformal
isometry algebra of the AdS$_3\times$S$^2$ background we consider
here. Such a dictionary was derived in \cite{deBoer:2008fk} and
can be straightforwardly applied to the \dihalo  case. The map
from supergravity to CFT quantum numbers is (recall that $c=6I$)
\be
L_0=N\,,\quad \tilde{L}_0=\frac{I}{4}\,,\quad J_3=-J\,.\label{Rcharges}
\ee
States with these quantum numbers are Ramond ground states, with
minimum eigenvalues under $\tilde L_0$, as expected for BPS
states. The calculation of the KK-spectrum on AdS$_3$, however, is
most naturally phrased in the NS sector and thus we would like to
work in this sector.  Thus we relate the charges (\ref{Rcharges})
by spectral flow \cite{Lerche:1989uy} in the right moving sector
to the charges of the corresponding states in the NS-sector.
Performing the spectral flow explicitly (see e.g
\cite{deBoer:2008fk} for some details) we find
\be
L_0=N\,,\quad \tilde{L}_0=\frac{I}{2}-J\,,\quad J_3=\frac{I}{2}-J\,.\label{NScharges}
\ee
As expected the BPS states manifest themselves in the NS sector as chiral
primaries, satisfying the condition $\bar L_0=J_3$. The well known unitarity
bound \cite{Lerche:1989uy} on the R-charge of chiral primaries implies a bound
on the range of the 4d angular momentum:
\be
0\leq J\leq \frac{I}{2}
\ee
From the results of the previous section it is clear that the
\dihalo configurations satisfy this bound. This bound was first
observed to have consequences for AdS$_3$/CFT$_2$ in
\cite{Maldacena:1998bw}, where it was called a stringy exclusion
principle. As was argued there, it has to be imposed by hand on
the free supergravity spectrum. What is perhaps surprising is that
in the fully interacting supergravity theory the bound seems to
emerge dynamically as it follows (at least for the \dihalo system)
from the integrability equations (\ref{d6d6d0_c1}) which are
essentially a consequence of the BPS equations of motion. We have no solid proof of this, but we were unable to find
other multicentered supergravity configurations that violate the
bound, even with flat space asymptotics where there is no direct
connection to the exclusion principle in the CFT. 

It is interesting to note that by (\ref{meaning}) and
(\ref{NScharges}) we see that for the \dihalo system
$\tilde{L}_0=M+1/2$ and that indeed also the bound on $M$,
observed in the previous section, follows directly from the
unitarity bound discussed above. Using the identification of $M$
and $\tilde{L}_0$, we can write the following analogue of the
generating function (\ref{partf}):
\be
\calz=\textrm{Tr}_{\mathrm{NS,BPS}}(-1)^F
q^{L_0}y^{\tilde{L}_0-1/2} .\label{cftpartf}
\ee
Some remarks are in order. First we would like to point out that,
for computational simplicity, we will calculate, in this section,
an index rather than an absolute number of states, the difference
with (\ref{partf}) being an explicit insertion of $(-1)^F$. As one
can see explicitly from the derivation below, the difference
between the absolute number of states and the index will be only
affect the numerical coefficient of the entropy, but not its
functional dependence on the charges. Second, note that at $y=0$
the above index coincides with the standard elliptic genus for
this theory.

\subsubsection{The spectrum of BPS states}

To calculate the degeneracies we are interested in, we need to enumerate the
possible BPS states of linearized (free) supergravity on AdS$_3\times$S$^2$.
It is often easier to enumerate these states via their quantum numbers in the
CFT so we will use this language.

As we only have supersymmetry in the right moving sector, there
are no BPS constraints on the left moving fields and thus all
descendants of highest weight states will appear. The right-moving
sector has $N=4$ supersymmetry and BPS states must be chiral
primaries of a given weight. As a consequence, and as was shown in
detail in e.g  \cite{Larsen:1998xm, deBoer:1998ip,
Fujii:1998tc, Kutasov:1998zh}, the full BPS spectrum
can be written in the form\footnote{Furthermore, to be fully
precise we should point out that there remain so called singleton
representations, but, for our purposes, we can ignore them as one
can show they only contribute to subleading terms in the entropy
in the large charge limit.}:
\be
\{s,\tilde h\}=\oplus_{n\geq0} \left(L_{-1}\right)^n|\tilde
h+s\rangle_L\otimes|\tilde h\rangle_R\label{tower}
\ee
where $|h\rangle_L$ are highest weight states of weight $h$ of the left-moving
Virasoro algebra and $|\tilde h\rangle_R$ are weight $\tilde h$ chiral primaries
of the right-moving $\caln=4$ super-Virasoro algebra.

Each field of five dimensional supergravity gives rise to a set of
BPS states and their descendants after KK-reduction, where $\tilde
h$ essentially labels the different spherical harmonics, while $n$
labels momentum excitations in AdS$_3$ and $s$ the spin of the
particle. It was shown in
\cite{Larsen:1998xm, deBoer:1998ip, Fujii:1998tc, Kutasov:1998zh}
that, given the precise field content of 5d $\caln=1$
supergravity, the reduction on a 2-sphere gives the set of quantum
numbers shown in table \ref{spectrum}.
\begin{table}
\begin{center}
\begin{tabular}{l|l|l}
5d origin & number & $\{s,\tilde h\}$-towers\\
\hline
    hypermultiplets & $2h^{1,2}+2$ & $\{\frac{1}{2},\frac{1}{2}+m\}$\\
    vectormultiplets & $h^{1,1}-1$ & $\{0,1+m\}$ and $\{1,m\}$ \\
    gravitymultiplet & $1$ & $\{-1,2+m\}$, $\{0,2+m\}$, $\{1,1+m\}$ and $\{2,1+m\}$
\end{tabular}\caption{Summary of the spectrum of chiral primaries on AdS$_3$.
The states are organized in towers of the form ({\protect
\ref{tower}}), the number of such towers and their characteristics
are determined by the properties of the original theory and the
details of the reduction. In the above table, $m$ is an arbitrary
nonnegative integer.} \label{spectrum}
\end{center}
\end{table}
Notice that the quantum numbers $\{s,\tilde{h}\}$ are of the form
$\{s,\tilde{h}_{\rm min} + m\}$, and for each such set the
partition function (\ref{cftpartf}) has the following form
\be
Z_{\{s,\tilde h_{\rm min}\}}=\prod_{n\geq0}\prod_{m\geq
0}(1-y^{m+\tilde h_{\rm min}-1/2}q^{n+m+\tilde h_{\rm
min}+s})^{(-1)^{2s+1}}
\ee
with the total partition function given by a product of such factors.

To extract the large $N$ degeneracies we proceed as in (\ref{freeenergy}) and
calculate the free energy corresponding to this partition function.  We then
evaluate it in the $\beta,\mu\ll1$ limit $(q=e^{-\b}, y=e^{-\m})$:
\bea
F_{\{s,h_{\rm min}\}}&=&(-1)^{2s}\sum_{n\geq1}\frac{q^{n(\tilde h_{\rm min}+s)}
y^{n \tilde h_{\rm min}}}{n (1-q^n)(1-y^nq^n)}\\
&\approx&\frac{(-1)^{2s}\zeta(3)}{\beta(\beta+\mu)}
\eea
Note that, as might have been expected, at high temperatures only
the statistics of the particles matter, as $h_{\rm min}$ and $s$
only change the lowest states of the towers. The total free energy
is now the sum over all different towers. Using table
\ref{spectrum} we find that
\be
F\approx \left[-(2h^{1,2}+2)+2(h^{1,1}-1)+4\right]
\frac{\zeta(3)}{\beta(\beta+\mu)}=\chi\frac{\zeta(3)}{\beta(\beta+\mu)}
\ee
where we used the definition of the Euler characteristic $\chi$ of the CY$_3$.
Finally we can do a Legendre transform to obtain the entropy.  This is
completely analogous to (\ref{legendre}) and the result is
\be
S\approx(\chi\zeta(3) M(N-M))^{1/3}
\ee
This result is equivalent to (\ref{legendre}) and maximization with respect to M
proceeds analogously, again leading to the result
\bea
S(N)=\begin{cases}\left(\chi\zeta(3)\frac{N^2}{4}\right)^{1/3}\,&\qquad\mbox{if}\
    N\leq I \label{entropy1}\\
\left(\chi\zeta(3)\,\frac{I}{2}(N-\frac{I}{2})\right)^{1/3}\,&\qquad\mbox{if}\ I\leq N
\end{cases}\label{ladida}
\eea
This might look somewhat unfamiliar when compared with other
calculations of the elliptic genus, e.g \cite{Gaiotto:2006ns, Kraus:2006wn}. This is because those calculations were all
performed in the regime $N\ll I$  where the
unitarity bound on the spectrum can be ignored. It is exactly
around $N\approx I$ that this bound starts to be relevant leading
to a different, slower, growth of the number of states in the
regime $I\ll N$. Such a behavior was also seen
in the computation of the elliptic genus in \cite{deBoer:1998us}.

Note that once more our computation above only applies for $N \lesssim I^2$ as
the asymptotic form of the free energy is essentially the same as that of the
dipole halo system.

\subsubsection{Comparison to black hole entropy and the stringy exclusion principle}

As we have seen, calculating the number of free supergravity states at fixed
total charge in the large $N, M$ limit proceeds rather analogously to the
counting of section \ref{sec_D6D6D0_entropy} and, more importantly, we found a
precise match between the leading contributions, up to an overall prefactor.

It is not hard, however, to see that even this prefactor can be made to match.
In the previous subsection we focussed on the 4 dimensional degrees of freedom
of the \dihalo-system ignoring the fact that the D0-branes bound to the
D6$\antiDp{6}$ still have degrees of freedom in the internal CY$_3$ manifold.
These internal degrees of freedom can be quantized via a 0+1 dimensional sigma
model\footnote{In this simplistic model we neglect more complicated interactions
coming from strings stretched between the D0's and the D6's in the CY.} on the
CY. The BPS states of this sigma model correspond to the cohomology of the
Calabi-Yau with even degree mapping to bosonic states and odd degree to
fermionic states.  Thus there are exactly $\chi$ BPS states per D0 when counted
with the correct sign, $(-1)^F$. Including this extra degeneracy in the
calculation of section \ref{sec_D6D6D0_entropy} will lead to a match with
(\ref{ladida}), including the prefactor.

That the two calculations provide the same amount of states is non-trivial,
since earlier we restricted ourselves to counting only states realized as a
\dihalo system, while in the second calculation we count all free supergravity
states in AdS$_3\times$S$^2$ with given momentum.  This suggests that indeed the
leading portion of supergravity entropy is realized as \dihalo configurations
once backreaction is included. This is a very strong result as clearly one can
think of many, much more complicated, smooth multicenter configurations with the
same total charge. Furthermore, we learn from these calculations that the number
of such states is parametrically smaller than the number of black hole states.
This seems to strongly indicate that the generic black hole state is associated
to degrees of freedom beyond supergravity.

From another perspective, however, the match between the free
regime and the \dihalo entropy is not so surprising.  If we
consider first a D6$\antiDp{6}$ bound state we can use a
coordinate transformation from \cite{Denef:2007yt} to map this to
global AdS$_3\times$S$^2$.  Thus we can think of the
D6$\antiDp{6}$ as simply generating the empty AdS background.
Recalling that D0 branes lift to gravitational shock waves in
5-dimensions one might already have anticipated that counting D0's
in the D6$\antiDp{6}$ background is closely related to counting
free gravitons on an AdS$_3\times$S$^2$ background. What makes the
result non-trivial is that interactions are apparently not
terribly relevant when counting BPS states, but then again
the D0's only interact very indirectly with each other. We might,
therefore, wonder if more exotic configurations, such as the
supereggs of \cite{Denef:2007yt} or the wiggling rings of
\cite{Bena:2008nh, Bena:2008dw}, are perhaps not captured by the
free theory and hence not subject to the bound we find above. The
problem with this is that we can compute not only the entropy but
also the index in both regimes and they exhibit the same leading
growth. If additional supergravity configurations are to generate
parametrically more states this would either require very precise
cancellations (so that the index is very different from the number
of states) or a phase transition at weak coupling (a phase
transition in $g_s$, not the $N=I$ transition discussed above).
Even if many states would cancel in the index, one would still
need to explain why they become invisible in the limit in which
interactions are turned off.

It is also somewhat intriguing to see that in the ``free theory''
we recover the phase transition noted in the previous section only
after imposing (by hand) the CFT unitarity bound suggesting that
the latter is taken into account by our scaling solutions. A
priori this sounds somewhat mysterious as the stringy exclusion
principle was argued in \cite{Maldacena:1998bw} to be inaccessible
to perturbative string theory.  As noted in \cite{deBoer:2008fk}
however the multicentered solutions seem to always satisfy this
bound (though there is no general proof of this).  In fact, the
origin of the bound in this system can simply be traced back to
the fact that the size of the angular momentum equals $j = I/2 -
M$, which cannot be negative, and using $M=L_0$ this then
immediately implies that the unitarity bound will be satisfied by
our solutions.

In the above, we have only counted multiparticle BPS
supergravitons in 5d supergravity. It is conceivable that
additional degrees of freedom could be obtained by allowing
fluctuations in the Calabi-Yau as well. For example, as we
discussed, D0-branes carry an extra degeneracy corresponding to
the harmonic forms on the Calabi-Yau. Though this can contribute a
finite multiplicative factor to the entropy, it does not change
the functional form. In addition, 5d supergravity does include all
massless degrees of freedom that one gets from the reduction on
the Calabi-Yau, and the other massive degrees of freedom
generically do not contain any BPS states.

One might also worry that multiparticle states, which in the free
theory are not BPS, become BPS once interactions are included.
Though this is a logical possibility, such degrees of freedom
would not contribute to the index, and therefore the estimate of
the index remains unaffected by this argument.

Finally, we notice that it is possible to do similar computations 
for AdS${}_3\times$S${}^3$, which leads to the result that for
$N\lesssim I$, $S\sim N^{3/4}$, while for $I \ll N  \ll I^2$, $S\sim I^{1/2}
N^{1/4}$. It would be interesting to reproduce these results by
counting solutions of 6d supergravity as well.

\section{Physical properties of scaling solutions}\label{sec_phys_scaling}

Although the counting above suggests that the class of \dihalo
solutions is not sufficiently generic to characterize macroscopic
five dimensional black holes the quantization has yielded several
interesting surprises that warrant further discussion.  Because of
the structure of scaling solutions it seems that the largest
number of states reside at the $J=0$ boundary of our solution
space once the charge $N$ carried by the D0 particles is
sufficiently large, i.e. $N>I$. Furthermore, as discussed in
\cite{deBoer:2008zn}, our quantization suggests a mass gap, for
asymptotically AdS$_3$ solutions, on the order of $1/c$ which is
somewhat unexpected from the perspective of the dual MSW
CFT\footnote{In some cases one can argue using dualities that the
MSW CFT admits a ``long string'' sector, in which case $1/c$ is
the natural value for the gap, similarly to the relation between
the F1-P and the D1-D5 system. For a generic MSW CFT we are not
aware of any duality which maps it into a system with an obvious
long string sector.}. Finally, in \cite{deBoer:2008zn} it was
found that a certain region of the D6$\antiDp{6}$D0 solution
space, corresponding to nearly coincident centers generating an
extremely deep throat, is characterized by an extremely low phase
space density implying that, upon quantization, it is not possible
to support classical states localized within this region.  In this
section we will explore some of these physically interesting
issues further.

\subsection{Scaling solutions with more than 3 centers}\label{sec-morethree}

A first generalization with respect to \cite{deBoer:2008zn} that we considered
in this paper are scaling solutions in which more than 3 centers are involved.
For example in figure \ref{4cpolytopes} we showed the different types of
polytopes for a dipole halo with 2 D0 centers. Depending on the distribution of
the D0 charge over those two centers there are now different types of polytopes
with a scaling region; cases B, C and D in figure \ref{4cpolytopes}. As in the
case of \cite{deBoer:2008zn} with only a single D0 center, the polytope is no
longer Delzant but rather rational when the scaling bound is saturated and the
scaling region is an orbifold singularity. Also, similar to the case with only a
single D0, the quantum wavefunctions have vanishing support on the locus
classically corresponding to coinciding centers.  As an example to illustrate
this feature we plot such a wavefunction in figure \ref{wavefunction}.

\FIGURE{\includegraphics[scale=0.7]{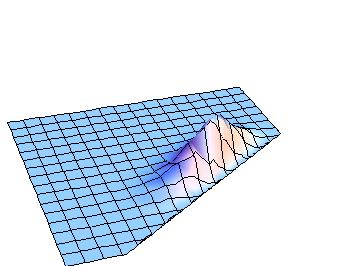}
\caption{For a dipole halo with two D0 branes of charge $q_1=10$, $q_2=14$ and
intersection product $I=32$ the wavefunction with quantum numbers $m_1=7$ and
$m_2=6$ is plotted as a function of the coordinates $y_1$ and $y_2$. The
dependence on $y$ has been removed by integration, as can be seen the
wavefunction has support on a polytope of the form B of figure {\protect
\ref{4cpolytopes}}. Note that this is one of a set of wavefunctions with zero
angular momentum as $J=I/2-3/2-m_1-m_2$. (A precise expression for the
wavefunctions in terms of given toric coordinates can be found in
\cite{deBoer:2008zn})\label{wavefunction} } }

In this figure the region where the centers coincide classically
in coordinate space, the ``scaling'' locus, is the diagonal edge
on the right, where $x_1+x_2=I/2$. While, this is a region
corresponding to zero angular momentum in the classical theory, in
the quantum theory there are intrinsic spin corrections to the
angular momentum (see figure \ref{spins} and the end of section
\ref{sec_phys_pic}) so the zero angular momentum states are
localized slightly away from the scaling locus. Put another way,
the intrinsic spin of the centers contributes a negative amount to
the total angular momentum which sets the quantum angular momentum
to zero before the classical $j=0$ point (which is the scaling
point) is reached.  Thus, as can be checked from the form of the
wavefunctions, there is zero probability to be at the classical
scaling point. As in \cite{deBoer:2008zn} we interpret this result
as an indication that the classical solutions near the scaling
point are highly quantum corrected, with fluctuations ranging over
a large space-time volume.

\subsection{Macroscopic quantum fluctuations from AdS/CFT}\label{sec_macro}

That the quantum mechanics of scaling solutions would necessarily
involve novel features was recognized shortly after their study in
the AdS/CFT context as it is in this context that they present the
most challenges.  The existence of infinitely smooth, deep
asymptotically AdS throats with low curvature everywhere seems to
suggest the existence of a continuous spectrum in the dual
conformal field theory.  This would not agree with the fact that
large black holes are dual to thermal states and at the same time
carry finite entropy. It would also disagree with our knowledge of
the spectrum of the D1-D5-P CFT at weak coupling.

As presaged in \cite[Section 6]{Bena:2007qc} resolving this would
require quantum effects that extend across large portions of
classical, smooth solutions and this is precisely along the lines
of what was found in \cite{deBoer:2008zn} by explicit computation.
The observation in \cite{deBoer:2008zn} that the phase space
volume of the system computed at weak coupling, for a system of
weakly interacting D-branes, does not increase as the branes
backreact and generate an infinitely deep throat follows naturally
from supersymmetry but is nonetheless remarkable as it implies
that the quantized ``cells'' of the BPS phase space stretch across
macroscopic volume as an infinite throat forms.  As such one might
worry that this property is somehow an artifact of the BPS nature
of the solutions. Here we would like to present, as supporting
evidence for this phenomenon, a generic AdS/CFT based argument
which uses some basic properties of the solution space, especially
the fact that it is a phase space, but which does not explicitly
rely on any supersymmetry.

To make an AdS/CFT based argument we of course require
asymptotically AdS solutions which can be obtained by taking an
appropriate decoupling limit of the solutions.  Thus we assume
that the total charge $\Gamma = \sum_a \Gamma_a$ has vanishing D6
charge, which allows us to take the decoupling limit of
\cite{deBoer:2008fk} and which allows us to generate a family of
asymptotically AdS$_3\times$S$^2$ solutions.

The essential observation is that generic harmonics in our solutions can be
expanded asymptotically as
\be
H = \sum_a \frac{\Gamma_a}{|\vec{x} - \vec{x}_b|} + h = \frac{\Gamma}{r} + h +
{\mathcal O}\biggl(\frac{|\vec{x}_a|}{r}\biggr)
\ee
where the terms of order zero in $x_a$ generate the base
AdS$_3\times$S$^2$ geometry (rather they generate the geometry of
an extremal BTZ black hole) and the subleading terms represent a
modification of this base geometry. AdS/CFT arguments relate the
expectation values of CFT operators in a particular state to
subleading terms in a boundary expansion of the geometry dual to
the state. In our solutions it is the terms proportional to $|\vec{x}_a|$
that generate these subleading terms in the expansion of the
fields.
For scaling solutions near the scaling point all the centers can
be arbitrarily close to the origin so $|\vec{x}_a| \sim \lambda
\ll 1$.  As $\lambda \rightarrow 0$ the solution develops an
infinitely deep scaling throat that closely resembles the naive
black hole geometry and solutions in this region all have
expectation values proportional to some positive power of
$\lambda$.

Because the solution spaces we are studying map to a symplectic
submanifold of the full phase space they contain both
configuration and conjugate momenta variables.  Hence we expect
that these solutions can be parameterized, in the dual theory, by
non-trivial expectation values of both an operator, $O$, and its
conjugate momentum conjugate, $\pi_O$.  Heisenberg's uncertainty
principle, however, implies there will be an intrinsic variance in
measuring these expectation values
\be
\sigma_O \sigma_{\pi_O} \geq 1
\ee
The crucial observation is that this bound on the variance is
finite and independent of $\lambda$ so as $\lambda \rightarrow 0$
(recall we are measuring expectation values in a state
$|\lambda\rangle$ dual to a throat parameterized by $\lambda$)
there will be some approximate value, $\lambda_c$, for which the
variance is of the same order as either $\langle O\rangle$ or
$\langle \pi_O \rangle$.  For such states $|\lambda\rangle$ we can
no longer think of the dual geometries as good classical solutions
as an observer doing measurements would not be able to distinguish
geometries corresponding to the different values of $\lambda$ and
these look macroscopically {\em very} different.


For instance, if we consider a dipole halo with only one D0 brane then from eqn.
(\ref{dipole_c1})-(\ref{deconst-ang-moment}) we see $\lambda \sim j \sim
\theta_a$ so the depth of the throat is controlled by the distance between the
D6 and the D0.  This can be measured in the CFT by measuring $\hat{J}_3$
(referred to as $J_0$ in \cite{deBoer:2008fk}).  The conjugate variable in the
bulk is $\sigma_a$ which parameterizes the phase space and also appears
asymptotically in the D0 dipole moment, $\vec{d}_0$ (see e.g.  \cite[eqn.
(3.29)]{deBoer:2008fk}).  For scaling solutions both of these asymptotic
coefficients will be first or higher order $\lambda$.  Thus both $\hat{J_3}$ and
its conjugate in the CFT will have expectation values and also variances of this
order (for small enough $\lambda$).  Even if we take the variance of $\langle
\hat{J}_3 \rangle$ to be very small in a state $|\lambda \rangle$ implying a
fixed throat depth the corresponding large variance in the dual operator implies
that the location of the D0 is smeared in a circle around the origin at the bottom
of the throat.  Recall that because of the warp factor the centers remain at a
fixed, macroscopic, distance apart so the throat ends in a large quantum foam
rather than a classical cap.

It would clearly be interesting to explore this argument in more
general cases, to make it more quantitative and to examine its
validity.

\subsection{Entropy of the $J=0$ locus} \label{sec_j0}

The quantization of $\mathcal{N}=2$ solutions in
\cite{deBoer:2008zn} resolved the paradox of infinitely deep
throats and the associated lack of a mass gap in the dual CFT by
an explicit computation.  A related question, posed in
\cite{Bena:2007qc}, is whether it is only the total angular
momentum $\vec{J}$ which is quantized, or whether each of the
``components'' of the angular momentum, $\vec{J}_{ij}$, are
separately quantized. The latter, combined with the boundedness of
these angular momenta, implies that the phase space has finite
volume, which in turn shows that infinitely deep throats
necessarily have to cut off.

While it is clear, from AdS/CFT, if nothing else, that $\vec{J}$ should be
quantized it is not clear that the $\vec{J}_{ij}$ should.  From the discussion
in section \ref{sec_phys_pic}, however, it is evident that our symplectic
form quantizes the angular momenta, $\vec{J}_{ij}$, between every pair of
centers $i, j$, not just the total angular momenta.

To pose the problem more sharply \cite{Bena:2007qc} suggests that one can
consider scaling solutions where the total angular momentum is zero (e.g. by
imposing $J=0$ as a boundary condition) and then ask if the same quantization
still works.  That the answer will be affirmative already follows, to a large
degree, from the discussion in section \ref{sec_phys_pic}.  Of course there are
some subtleties that must be addressed.  The $j$ term in the symplectic form,
related to {\em overall} rotations of the system (this is clear in the
coordinate system of section \ref{deconstruct_setup}) depends on the orientation
of $\vec{J}_{6a}$.  On the other hand setting $j$ to zero\footnote{Note the
subtlety involving the difference between the locations where the angular
momentum vanishes classically and where it vanishes when intrinsic quantum spin
effects are included, as explained in section \ref{sec-morethree}.} actually
imposes the least restriction on the range of the $j_{a} \cos \theta_a$.  This
follows immediately from eqn.  (\ref{scalingcond}) and implies that, at least in
the scaling regime, a large number of states sit at the $j=0$ locus.  This is
can be directly seen from figure \ref{4cpolytopes}, e.g. case D, where the $j=0$
locus is the diagonal edge of the $y=0$ subspace (the 2d polytope on the left.)

Somewhat more formally this $j=0$ submanifold is given in the polytope as the
subspace $\{y=0, y_1, \dots, y_n\}$ with $\sum_a y_a = I/2$ and one can
explicitly check that this is a symplectic submanifold of the large phase space
on which the  pull-back of the symplectic form is non-degenerate. For instance,
in the case with two D0 centers it corresponds to the quantization of the
difference $\vec{J}_{61}-\vec{J}_{62}$.

Finally, as an additional argument we quote a mathematical result of
\cite{Guillemin:1982}.  Namely, (geometric) quantization of the fixed point set
of the Hamiltonian action of a compact group on a phase space  yields a Hilbert
space isomorphic to the invariant (under the induced group action) subspace of
the Hilbert space obtained by quantizing the entire phase space.

In this case the group SO(3) (or its cover, SU(2)), corresponding to ${\mathbb R}^3$
rotations, acts on the phase space and the theorem implies that rather than restrict
to the $j=0$ subspace and quantize that anew we can simply use our existing
quantization and extract the invariant subspace of our Hilbert space.  From
figure \ref{4cpolytopes}, cases B, C and D, it is then clear that the $j=0$ locus contributes a significant
number of states.

\section*{Acknowledgments}

We would like to thank I.~Bena, M.~Berkooz, F.~Denef, E.~Diaconescu, S.~Guisto,
J.~Manschot, J.~Maldacena, G.~Moore and C.~Ruef for interesting and clarifying discussions.

The work of JdB, SES and IM is supported financially by the Foundation of
Fundamental Research on Matter (FOM). The work of DVdB is supported by the DOE under grant DE-FG02-96ER40949.

\appendix

\section{The number of states as discrete points inside the polytope}\label{diet-proof}

In this appendix we show that for {\em rational} polytopes the number of
normalizable modes can simply be computed from the 'discretized volume' of the
polytope.  This can be useful as it saves time for problems where we are only
interested in the number of states and not in the explicit wave functions.

Let us very shortly review the description we use for polytopes,
for the full definition and the algorithms and formulae to
calculate the associated complex coordinates and wavefunctions we
refer the reader to appendix B of \cite{deBoer:2008zn}.

Basically, we can think of a toric polytope as a region in
$\mathbb{R}^n$, parameterized by coordinates $x^i$, on which a
certain set of first order polynomials are positive. That is,
given such a set of $m$ first order functions:
\be \label{defpoly}
l_j(x)=\sum_{i=1}^nc_{ij}x^i+\lambda_j\,
\ee
the polytope is defined as $P_{l}=\{x\in\mathbb{R}^n|l_j(x)\geq0\}$. A few
remarks are in order:
\begin{itemize}
\item  It follows from this definition that the polytope is an intersection of
    $m$ half spaces.
\item Note that from that interpretation it follows that $m\geq2+n$ to have a compact
polytope, so $c_{ij}$ is never a square matrix! (this has some consequences later)
\item Note that $c_{ij}$ here is actually the transpose of the one defined in
    the appendix of \cite{deBoer:2008zn}.  This because we now use the more
    natural definition $c_{ij}=\frac{\partial l_j}{\partial x^i}$.
\item Finally note that $c$ and $\lambda$ cannot be completely arbitrary (i.e.
    not every intersection of half planes gives a sensible polytope).
\end{itemize}

\paragraph*{Example} For the readers convenience we will give the defining
functions corresponding to the dipole halos of equations
(\ref{standcond})-(\ref{scalingcond}).  Let us first define a
coordinate system using the $n+1$ coordinates $(y_0, y_1, \dots,
y_n)$ corresponding to (\ref{ycoords}) with $y_0 = y$..  To encode
these constraints in a polytope we require an $(n+1)\times(2n+3)$
$c_{ij}$ matrix which which we will think of instead as $2n+3$
vectors of length $n+1$ given below. In addition we will also need
a $2n+3$-component vector $\lambda$ with components given below as
well.
\begin{align}
\vec{c}_{0} &= (-1, \dots, -1) \qquad &\lambda_0 &= I/2 \nonumber \\
\vec{c}_{1} &= (1, -1, \dots, 1) \qquad &\lambda_1 &= I/2 \nonumber \\
\vec{c}_{2a} &= (0, \dots, -1, \dots, 0) \qquad &\lambda_{2a} &= q_a
\label{defscalepoly} \\
\vec{c}_{2a+1} &= (0, \dots, 1, \dots, 0) \qquad &\lambda_{2a+1} &= 0 \nonumber \\
\vec{c}_{2n+2} &= (0, -1, \dots, -1) \qquad &\lambda_{2n+2} &= I/2 \nonumber
\end{align}
The $\vec{c}_{2a}$ and $\vec{c}_{2a+1}$ are non-zero only on the $(a+1)$'th
entry (recall that $a=1,\dots,n$ and our coordinates are labelled from
$0,\dots,n$).  Note that the indices on $\vec{c}$ correspond to the labels $j$
in (\ref{defpoly}).  With this in mind the reader can check that the $2n+3$
equations defined by substituting the $\vec{c}$ and $\lambda$ above into
(\ref{defpoly}) reproduce (\ref{standcond})-(\ref{scalingcond}).
The corresponding polytopes are shown in figure \ref{4cpolytopes}.\\
\\

As discussed in the Appendix B of \cite{deBoer:2008zn} and
references therein, all relevant functions (i.e. complex
coordinates, K\"ahler potential, etc.) are defined in terms of the
$c$ and $\lambda$. Hence we can write the norm square of the
wavefunction $\psi_a=\prod_i(z^i)^{(a^i)}$, with
$a\in\mathbb{Z}^n$ and the $z_i$ appropriate complex coordinates
on the toric manifold, in terms of these objects:
\bea
|\psi_a|^2\sim e^{\sum_i\partial_i g}\sqrt{\det\partial_i\partial_j g}\,e^{-\calk}\prod_{i}|z^i|^{(2a^i)}&\sim& e^{\sum_i\partial_i g}\sqrt{\det\partial_i\partial_j g}\prod_{j=1}^{m}l_j^{(\sum_{i=1}^n c_{ij}a^i+\lambda_j)}\label{step1}\\&\sim&\prod_{j=1}^ml_j^{\left(\sum_{i=1}^nc_{ij}(a_i+1/2)+\lambda_j-1/2\right)}\label{poles}
\eea
where $\sim$ indicates proportionality up to constants and functions that have no
poles and also contain no overall $l_j$ factors. The first step is rather
straightforward while the last step is more subtle to prove so that proof is
relegated to a separate subsection below.

Let us first focus on the interpretation of the above result. We
see that, without taking into account the fermionic contribution
$e^{\sum_i\partial_i g}\sqrt{\det\partial_i\partial_j g}$,
normalizability of the wavefunctions requires the
$a\in\mathbb{Z}^{n}$ to satisfy
\be
\sum_{i=1}^n c_{ij}a^i+\lambda_j>-1
\ee
while, when also including those necessary fermionic corrections,
we find the final precise condition is
\be
\sum_{i=1}^nc_{ij}(a_i+1/2)+\lambda_j-1/2>-1\label{states=volume}
\ee
Up to some shifts these equations essentially tell us that $a$ has
to lie ``inside'' the polytope, making the number of states
essentially the discretized volume of the polytope, i.e. the
volume dived in Planck size cells. Furthermore, as we discuss in
detail in the specific case studied in the main text, the shifts
of 1/2 introduced by taking into account the fermionic nature of
the wavefunctions has a very natural physical interpretation. As
was discussed in \cite{deBoer:2008zn} the quantization of the
polytopes roughly corresponds to quantizing the angular momentum
of the system.  That the lowest energy state corresponds to a
specific alignment of the spins of the constituents then leads to
various half integer shifts of the quantum angular momentum,
giving rise to the 1/2's in (\ref{states=volume}).

Plugging the $\vec{c}$'s and $\lambda$'s defined in (\ref{defscalepoly}) into
eqn. (\ref{states=volume}) should allow the reader to reproduce the constraints
found in (\ref{nonorbcond}).

\subsection{Evaluation of $\det\partial_i\partial_j g$}
In this subsection we give a detailed description of the steps
which lead from (\ref{step1}) to (\ref{poles}). These steps are
based on an intermediate result, which states that
\be
\det\partial_i\partial_jg=\left(\prod_{j=1}^{m}\frac{1}{l_j}\right)A(l)\label{detform}\,,
\ee
where as we will show $A(l)$ is a homogeneous polynomial of order
$m-n$ in the $l_j$ with such coefficients that for no rational
polytopes it will contain an overall $l_j$ factor.

We will prove this in two steps. First we will evaluate the
relevant determinant to show the form (\ref{detform}) explicitly.
In the second step we then use this explicit form of $A(l)$ to
argue its relevant properties, namely that it has no poles nor
contains an overall $l_j$ factor.
\paragraph{Calculating the determinant}
It is straightforward to check that \be
\partial_i\partial_j g=\frac{1}{2}\sum_{k=1}^{m}\frac{c_{ik}c_{jk}}{l_k}=\frac{1}{2}\left(C\cdot L^{-1}\cdot C^T\right)_{ij}
\ee
with $C_{ij}=c_{ij}$, remember $\dim C=n\times m$, and
$L_{ij}=l_j\delta_{ij}$ an $m\times m$ matrix. So indeed
$CL^{-1}C^T$ is a square $n\times n$ matrix and the determinant
makes sense, sadly the factors inside are not square matrices
making the evaluation a bit less straightforward. (We are not
interested in constant factors so we will forget about the 1/2 in
the following)

Using the basic definition of the determinant and using some
symmetry properties it is not too difficult, though maybe a bit
tedious to show that
\be
\det\left(CL^{-1}C^T\right)=\left(\prod_{j=1}^ml_j^{-1}\right)\left(\sum_S l^S(\det C_S)^2\right)\,.\label{detformula}
\ee
The second factor might need some explanation as it uses some
unconventional notation. The sum is over all different subsets
$S\subset\{1,\ldots,m\}$ with $m-n$ elements, i.e. $\# S=m-n$.
Furthermore we use the shorthand $l^S:=\prod_{j\in S}l_j$. Finally
there is the definition of the $n\times n$ matrix $C_S$. Note that
$C$ was an $n\times m$ matrix, $C_S$ is now defined as the matrix
$C$ but with the $i_1,\ldots, i_{m-n}$'th columns removed where
$S=\{i_1,\ldots,i_{m-n}\}$.

\paragraph{Properties of $A(l)$}
We found the result of (\ref{detform}) with the explicit form
$A(l)=\sum_S l^S(\det C_S)^2$. We now want to show that $A(l)$ has
no poles nor that it contains an overall $l_j$ factor. As is clear
from its definition $A(l)$ is a homogeneous polynomial of order
$m-n$ in the $l_j$. As the $l_j$ themselves are simply first order
in the $x^i$, the polynomial $A$ has no poles in the $x^i$. The
second point, that there is no overall $l_j$ factor, is more
subtle to see. To show it, pick a particular element
$j\in{1,\ldots,m}$. By relabeling we can just take $j=1$. Now from
its definition it is clear that $A(l)$ only has an overall $l_1$
factor if the coefficients of all the terms $l^{\tilde S}$, with
$\tilde S$ such that $1\notin \tilde S$, vanish.

We can now easily show that this never happens using some basic
properties of $C$ and $C_{\tilde S}$. We will argue that there is
always at least one $\tilde S_\star$ among the $\tilde S$ for
which $\det C_{\tilde S_\star}$ doesn't vanish. By the definition
of the $C_S$, all the $C_{\tilde S}$ include the first column of
$C$, given by $c_{i1}$.  Furthermore let us go  go back to the
definition of $C$ and the $c_{ij}$. Note that the original
definition of $c_{ij}$ was that it consisted of the $n$ components
of the $\vec{c}_{j}$, which were the normals to the $m$ facets of
the polytope. The statement $\exists \tilde S_\star\,|\,\det
C_{\tilde S_\star}\neq0$ thus translates to: "there exists a set
of $(n-1)$ vectors among the $m$ different normals $\vec{c}_j$
that together with $\vec{c}_1$ form a basis of $\mathbb{R}^n$\,".
We will use the notation $\vec{c}_a$ for these $n$ vectors and now
show their existence.

Pick one of the vertices that is a corner of the facet orthogonal
to $\vec{c}_1$ and let's call it $v_1$. As the polytopes of our
interest are rational there are exactly $n$ edges $\vec{e}_i$
meeting in the vertex $v_1$, that furthermore form a basis of
$\mathbb{R}^n$. Now the different facets meeting in $v_1$ each lie
in a subspace generated by a set of $(n-1)$ of the $n$ edges
$e_i$\ \footnote{Note that all facets are of this form by the
definition of the edges. That furthermore each of the $n$
combinations of $n-1$ linearly independent edges generates a facet
is maybe less straightforward and actually not true for a generic
non-rational polytope. However here the fact that for each
subspace generated by ($n$-1) of the $n$-edges there is only one
remaining edge not contained in that subset, implies that the
subspace must be on the boundary of the polytope and hence
generate a facet.}. So we find $n$ facets that all meet in the
vertex $v_1$. Let us label the $n$ normals to these facets as
$\vec{c}_a$, by their definition they can be labelled such that
they satisfy $\vec{e}_i\cdot \vec{c}_a\sim\delta_{i,j}$. So we see
that the $\vec{c}_a$ form a basis of $\mathbb{R}^n$ that includes
$\vec{c}_1$, which concludes the proof, i.e. we now know that
$\det C_{\tilde S_\star}\neq0$ for $(C_{\tilde
S_\star})_{ia}=c_{ia}$.

\bibliographystyle{JHEP}
\bibliography{refslist-proc}

\providecommand{\href}[2]{#2}\begingroup\raggedright\begin{thebibliography}{10}

\bibitem{Maldacena:1997de}
J.~M. Maldacena, A.~Strominger, and E.~Witten, {\it Black hole entropy in
  m-theory},  {\em JHEP} {\bf 12} (1997) 002,
  [\href{http://xxx.lanl.gov/abs/hep-th/9711053}{{\tt hep-th/9711053}}].

\bibitem{deBoer:2008zn}
J.~de~Boer, S.~El-Showk, I.~Messamah, and D.~Van~den Bleeken, {\it {Quantizing
  N=2 Multicenter Solutions}},  \href{http://xxx.lanl.gov/abs/0807.4556}{{\tt
  arXiv:0807.4556}}.

\bibitem{Denef:2000ar}
F.~Denef, {\it {On the correspondence between D-branes and stationary
  supergravity solutions of type II Calabi-Yau compactifications}},
  \href{http://xxx.lanl.gov/abs/hep-th/0010222}{{\tt hep-th/0010222}}.

\bibitem{Bates:2003vx}
B.~Bates and F.~Denef, {\it Exact solutions for supersymmetric stationary black
  hole composites},  \href{http://xxx.lanl.gov/abs/hep-th/0304094}{{\tt
  hep-th/0304094}}.

\bibitem{Bena:2005va}
I.~Bena and N.~P. Warner, {\it Bubbling supertubes and foaming black holes},
  {\em Phys. Rev.} {\bf D74} (2006) 066001,
  [\href{http://xxx.lanl.gov/abs/hep-th/0505166}{{\tt hep-th/0505166}}].

\bibitem{Berglund:2005vb}
P.~Berglund, E.~G. Gimon, and T.~S. Levi, {\it Supergravity microstates for bps
  black holes and black rings},  {\em JHEP} {\bf 06} (2006) 007,
  [\href{http://xxx.lanl.gov/abs/hep-th/0505167}{{\tt hep-th/0505167}}].

\bibitem{deBoer:2008fk}
J.~de~Boer, F.~Denef, S.~El-Showk, I.~Messamah, and D.~Van~den Bleeken, {\it
  {Black hole bound states in AdS$_3$ x S$^2$}},  {\em JHEP} {\bf 11} (2008)
  050, [\href{http://xxx.lanl.gov/abs/0802.2257}{{\tt arXiv:0802.2257}}].

\bibitem{Maldacena:1998bw}
J.~M. Maldacena and A.~Strominger, {\it Ads(3) black holes and a stringy
  exclusion principle},  {\em JHEP} {\bf 12} (1998) 005,
  [\href{http://xxx.lanl.gov/abs/hep-th/9804085}{{\tt hep-th/9804085}}].

\bibitem{Mathur:2005zp}
S.~D. Mathur, {\it The fuzzball proposal for black holes: An elementary
  review},  {\em Fortsch. Phys.} {\bf 53} (2005) 793--827,
  [\href{http://xxx.lanl.gov/abs/hep-th/0502050}{{\tt hep-th/0502050}}].

\bibitem{Bena:2007kg}
I.~Bena and N.~P. Warner, {\it Black holes, black rings and their microstates},
   \href{http://xxx.lanl.gov/abs/hep-th/0701216}{{\tt hep-th/0701216}}.

\bibitem{Skenderis:2008qn}
K.~Skenderis and M.~Taylor, {\it {The fuzzball proposal for black holes}},
  \href{http://xxx.lanl.gov/abs/0804.0552}{{\tt arXiv:0804.0552}}.

\bibitem{Mathur:2008nj}
S.~D. Mathur, {\it {Fuzzballs and the information paradox: a summary and
  conjectures}},  \href{http://xxx.lanl.gov/abs/0810.4525}{{\tt
  arXiv:0810.4525}}.

\bibitem{Bena:2008nh}
I.~Bena, N.~Bobev, C.~Ruef, and N.~P. Warner, {\it {Entropy Enhancement and
  Black Hole Microstates}},  \href{http://xxx.lanl.gov/abs/0804.4487}{{\tt
  arXiv:0804.4487}}.

\bibitem{Balasubramanian:2008da}
V.~Balasubramanian, J.~de~Boer, S.~El-Showk, and I.~Messamah, {\it {Black Holes
  as Effective Geometries}},  {\em Class. Quant. Grav.} {\bf 25} (2008) 214004,
  [\href{http://xxx.lanl.gov/abs/0811.0263}{{\tt arXiv:0811.0263}}].

\bibitem{Gaiotto:2004ij}
D.~Gaiotto, A.~Strominger, and X.~Yin, {\it {Superconformal black hole quantum
  mechanics}},  {\em JHEP} {\bf 11} (2005) 017,
  [\href{http://xxx.lanl.gov/abs/hep-th/0412322}{{\tt hep-th/0412322}}].

\bibitem{Kim:2005yb}
S.~Kim and J.~Raeymaekers, {\it {Superconformal quantum mechanics of small
  black holes}},  {\em JHEP} {\bf 08} (2005) 082,
  [\href{http://xxx.lanl.gov/abs/hep-th/0505176}{{\tt hep-th/0505176}}].

\bibitem{Denef:2007yt}
F.~Denef, D.~Gaiotto, A.~Strominger, D.~Van~den Bleeken, and X.~Yin, {\it Black
  hole deconstruction},  \href{http://xxx.lanl.gov/abs/hep-th/0703252}{{\tt
  hep-th/0703252}}.

\bibitem{Gimon:2007mha}
E.~G. Gimon and T.~S. Levi, {\it {Black Ring Deconstruction}},
  \href{http://xxx.lanl.gov/abs/0706.3394}{{\tt 0706.3394}}.

\bibitem{Levi:xxx}
T.~Levi, J.~Raeymaekers, D.~Van~den Bleeken, W.~Van~Herck, B.~Vercnocke, and
  T.~Wyder, {\it To appear}, .

\bibitem{Bena:2006kb}
I.~Bena, C.-W. Wang, and N.~P. Warner, {\it {Mergers and typical black hole
  microstates}},  {\em JHEP} {\bf 11} (2006) 042,
  [\href{http://xxx.lanl.gov/abs/hep-th/0608217}{{\tt hep-th/0608217}}].

\bibitem{Denef:2007vg}
F.~Denef and G.~W. Moore, {\it Split states, entropy enigmas, holes and halos},
   \href{http://xxx.lanl.gov/abs/hep-th/0702146}{{\tt hep-th/0702146}}.

\bibitem{Bena:2007qc}
I.~Bena, C.-W. Wang, and N.~P. Warner, {\it Plumbing the abyss: Black ring
  microstates},  \href{http://xxx.lanl.gov/abs/0706.3786}{{\tt 0706.3786}}.

\bibitem{Denef:2000nb}
F.~Denef, {\it Supergravity flows and d-brane stability},  {\em JHEP} {\bf 08}
  (2000) 050, [\href{http://xxx.lanl.gov/abs/hep-th/0005049}{{\tt
  hep-th/0005049}}].

\bibitem{Denef:2002ru}
F.~Denef, {\it Quantum quivers and hall/hole halos},  {\em JHEP} {\bf 10}
  (2002) 023, [\href{http://xxx.lanl.gov/abs/hep-th/0206072}{{\tt
  hep-th/0206072}}].

\bibitem{Raeymaekers:2007ga}
J.~Raeymaekers, {\it {Near-horizon microstates of the D1-D5-P black hole}},
  {\em JHEP} {\bf 02} (2008) 006,
  [\href{http://xxx.lanl.gov/abs/0710.4912}{{\tt arXiv:0710.4912}}].

\bibitem{Bena:2008dw}
I.~Bena, N.~Bobev, C.~Ruef, and N.~P. Warner, {\it {Supertubes in Bubbling
  Backgrounds: Born-Infeld Meets Supergravity}},
  \href{http://xxx.lanl.gov/abs/0812.2942}{{\tt arXiv:0812.2942}}.

\bibitem{Balasubramanian:2006gi}
V.~Balasubramanian, E.~G. Gimon, and T.~S. Levi, {\it {Four Dimensional Black
  Hole Microstates: From D-branes to Spacetime Foam}},  {\em JHEP} {\bf 01}
  (2008) 056, [\href{http://xxx.lanl.gov/abs/hep-th/0606118}{{\tt
  hep-th/0606118}}].

\bibitem{Larsen:1998xm}
F.~Larsen, {\it {The perturbation spectrum of black holes in N = 8
  supergravity}},  {\em Nucl. Phys.} {\bf B536} (1998) 258--278,
  [\href{http://xxx.lanl.gov/abs/hep-th/9805208}{{\tt hep-th/9805208}}].

\bibitem{deBoer:1998ip}
J.~de~Boer, {\it {Six-dimensional supergravity on S**3 x AdS(3) and 2d
  conformal field theory}},  {\em Nucl. Phys.} {\bf B548} (1999) 139--166,
  [\href{http://xxx.lanl.gov/abs/hep-th/9806104}{{\tt hep-th/9806104}}].

\bibitem{Lerche:1989uy}
W.~Lerche, C.~Vafa, and N.~P. Warner, {\it {Chiral Rings in N=2 Superconformal
  Theories}},  {\em Nucl. Phys.} {\bf B324} (1989) 427.

\bibitem{Fujii:1998tc}
A.~Fujii, R.~Kemmoku, and S.~Mizoguchi, {\it {D = 5 simple supergravity on
  AdS(3) x S(2) and N = 4 superconformal field theory}},  {\em Nucl. Phys.}
  {\bf B574} (2000) 691--718,
  [\href{http://xxx.lanl.gov/abs/hep-th/9811147}{{\tt hep-th/9811147}}].

\bibitem{Kutasov:1998zh}
D.~Kutasov, F.~Larsen, and R.~G. Leigh, {\it {String theory in magnetic
  monopole backgrounds}},  {\em Nucl. Phys.} {\bf B550} (1999) 183--213,
  [\href{http://xxx.lanl.gov/abs/hep-th/9812027}{{\tt hep-th/9812027}}].

\bibitem{Gaiotto:2006ns}
D.~Gaiotto, A.~Strominger, and X.~Yin, {\it From ads(3)/cft(2) to black holes /
  topological strings},  \href{http://xxx.lanl.gov/abs/hep-th/0602046}{{\tt
  hep-th/0602046}}.

\bibitem{Kraus:2006wn}
P.~Kraus, {\it Lectures on black holes and the ads(3)/cft(2) correspondence},
  \href{http://xxx.lanl.gov/abs/hep-th/0609074}{{\tt hep-th/0609074}}.

\bibitem{deBoer:1998us}
J.~de~Boer, {\it {Large N Elliptic Genus and AdS/CFT Correspondence}},  {\em
  JHEP} {\bf 05} (1999) 017,
  [\href{http://xxx.lanl.gov/abs/hep-th/9812240}{{\tt hep-th/9812240}}].

\bibitem{Guillemin:1982}
V.~Guillemin and S.~Sternberg, {\it Geometric quantization and multiplicities
  of group representations},  {\em Inventiones Mathematicae} {\bf 67} (1982)
  515--538.

\end{thebibliography}\endgroup

\end{document}